# The Long-Term Effects of Teachers' Gender Stereotypes [*]

Joan J. Martínez

UC Berkeley

Click here for the most updated version

**Abstract**

This paper studies the effects of teachers' stereotypical assessments of boys and girls on students' long-term outcomes, including high school graduation, college attendance, and formal sector employment. I measure teachers' gender stereotypical grading based on preconceptions about boys' aptitude in math and science and girls' aptitude in communicating and verbal skills by analyzing novel data on teachers' responses to the Implicit Association Test (IAT) and differences in gender gaps between teacher-assigned and blindly graded tests. To collect IAT scores on a national scale, I developed a large-scale educational program accessible to teachers and students in Peruvian public schools. This analysis provides evidence that teachers' gender stereotypes are reflected in student evaluations: math teachers with stronger stereotypes associating boys with scientific disciplines award male students with higher test scores compared to blindly-graded test scores. In contrast, language arts teachers who stereotypically associate females with humanities-based disciplines give female students higher grades. Using graduation, college enrollment, and matched employer-employee data on 1.7 million public high school students expected to graduate between 2015 and 2019, I find that female students who are placed with teachers who use more stereotypical grading practices are less likely to graduate from high school and apply to college than male students. In comparison to their male counterparts, female high school graduates whose teachers employ more stereotypical grading practices are less likely to be employed in the formal sector and have fewer paid working hours. Furthermore, exposure to teachers with more stereotypical grading practices reduces women's monthly earnings, thereby widening the gender pay gap.

[*]I am indebted to Christopher Walters, Patrick Kline and Jesse Rothstein for their extensive support and guidance. I am thankful for comments from Sydnee Caldwell, Christopher Campos, David Card, Michela Carlana, Supreet Kaur, Camille Landais, Lukas Leucht, Edward Miguel, Enricco Moretti, Amanda Pallais, and Damian Vergara. I also thank seminar participants Chicago Harris School of Public Policy, Bocconi University, Boston University, Harvard Kennedy School, the London School of Economics, Ohio State, Clemson University, the UC Berkeley Labor Seminar, and the Development Seminar. I'm thankful for the institutional support and information access from the Ministry of Education and the Ministry of Labor of Peru. I thank Ayrton Dextre, Daniela Rossini, and Jose Miguel Valverde for assistance in the data collection. I gratefully acknowledge financial support from the UC Berkeley Economics Department, the UC Berkeley Opportunity Lab, the Weiss Fund for Research in Development Economics, the Institute for Research on Labor Employment Dissertation Fellowship, and the Center for Effective Global Action through its Psychology and Economics of Poverty Initiative and Gender-Related Research in Economics initiative. IRB approval was received from University of California, Berkeley. First version: October 30, 2022. This version: May 8, 2023. Corresponding author: martinezp_jj@berkeley.edu

# 1   Introduction

Although gender wage gaps have declined worldwide in recent decades, this pattern has been uneven between developed economies (on the one side) and middle- and low-income countries (on the other), where gaps are still substantial. Moreover, labor supply outcomes continue to differ across genders globally, as female workers are more likely to experience unemployment and part-time employment, and female representation in high-paying industries and occupations remains uneven.[1] Recent evidence suggests that gender differences in salary requests (Rousille, 2021), self-promotion (Exley and Kessler, 2022), job-attribute preferences (Le Barbanchon et al., 2020, Biasi and Sarsons, 2020, Fleche et al., 2018, Flory et al., 2015), competitiveness (Niederle and Vesterlund, 2007), and willingness to contribute in team settings (Coffman, 2014), contribute to these gaps. These differences may emerge in the workplace as a result of gendered social norms and stereotypes that workers encounter early on in life and at school.

In this article, I focus on the role of educators' stereotyped assessments in generating gender differences in performance and education decisions, which have repercussions for future labor market outcomes. Numerous studies have demonstrated that educators have a lasting impact on their students' careers long after they leave the classroom (Chetty et al., 2014a,b, Rothstein, 2010, 2017). Nevertheless, evidence on the impact of teachers' stereotyped assessments is necessary for disentangling the long-term effects of value-added and narrowing the performance gap between student groups. It has been hypothesized that teachers' subjective views and preconceptions about boys' science and math talent and girls' verbal and communicating skills, which might be unrelated to students' actual academic performance, influence their effort allocation, resource allocation, evaluations, and encouragement of students across knowledge domains. Recent studies provide evidence that prejudices and stereotypes among educators discourage girls from pursuing science-focused high school tracks (Alesina et al., 2018, Carlana, 2019), increase gender score gaps (Lavy, 2008, Lavy and Sand, 2018), and influence short- and medium-term career decisions (Lavy and Megalokonomou, 2019). However, the longer-term effects of stereotyped evaluations on the labor market have been more difficult to assess for two primary reasons. First, there is an ongoing debate about the accuracy of measuring gender prejudices and stereotypes, and second, there has been a dearth of information on students' professional trajectories in labor markets long after exposure to stereotyped teachers. I aim to contribute to the existing literature by filling this gap.

This paper studies the extent to which gender stereotyped teacher assessments of student performance affect their students' educational careers, college applications, college attendance, and labor market outcomes. I use novel data from two sources: linked administrative information containing students' academic and professional trajectories from ages 12 to 23 and nationwide survey responses from teachers and students in Peru's public high schools. Two measures, systematic assessment gaps between boys and girls and the Implicit Association Test (IAT), have been at the center of debate regarding their ability to measure implicit and explicit stereotypes influencing female and male students' assessments in math and science relative to communication and the humanities. The IAT is a psychological test based on the differences in response times between associating boys with math-science disciplines and girls with humanities disciplines (Greenwald et al., 2003, 1998, Nosek et al., 2007, 2014). Stronger implicit preferences for the concepts are revealed by faster associations between pairs. I construct both of these measures by using students' test scores on two types of examinations (i.e., teacher- and blindly-graded) and by developing a government educational program through which I recorded approximately 2,400 teacher responses and 4,600 student responses to the Implicit Association Test and gender-attitudes surveys.[2] In addition, by combining comprehensive administrative records on students' education with their employment records, I am able to evaluate the long-term effects of exposure to stereotypical performance assessments on the pathways of five cohorts of students from high school to the formal labor market. My identification strategy compares students enrolled in the same cohort-grade-year school cells but assigned to different teachers, and consequently subjected to varying degrees of stereotypical assessment practices.

---

[1] According to reports by OECD (2022), ILO (2018), and ILO (2019), the gender wage gap in OECD countries today is 11.7 percent, while in middle- and low-income countries, it remains higher at 15–27 percent. ILO (2020) and World Economic Forum (2018) report labor-supply stylized facts.

My analysis begins with the estimation of an individual-level measure for gender differences in student evaluations based on disparities in gender gaps between teacher-assigned and blindly graded assessments (Botelho et al., 2015, Burgess and Greaves, 2013, Carrell et al., 2010, Lavy, 2008, Lavy and Sand, 2018). Using this metric, I am able to identify teachers who, relative to other teachers in the same subjects, are more likely to assign high grades to boys than to girls (or the converse) in classroom examinations relative to blindly graded tests in mathematics, science, and language arts. I control for a comprehensive set of observable teacher and student characteristics (e.g., place of birth, age, cognitive abilities proxy) in order to exclude the possibility that discrepancies in assessments are caused by any other observable source related to students' gender besides teachers' stereotypes. I interpret the remaining systematic gender differences in favor of boys (or girls) in classroom evaluations as reflecting teachers' stereotypical views and preconceptions of students' abilities in numeric-scientific and communication tasks. Using newly collected data from the Implicit Association Test (IAT) on a subsample of teachers for whom I have information on gender-math stereotypes, I test this hypothesis.

To the best of my knowledge, this is the first study to demonstrate that individual gender-math stereotypes robustly predict observed behavior, as evidenced by the gender gaps in teachers' assessments relative to centrally assigned scores. I provide evidence that the average Implicit Association Test score of mathematics teachers is 0.301, indicating strong associations between boys and math-science and between girls and the humanities. These estimates indicate that girls in developing countries are more likely to encounter teachers who hold negative stereotypes about their adequacy at solving math problems.[3] Importantly, the direction of the relationship between the Implicit Association Test scores and the estimates of teacher stereotyped assessment provides compelling evidence that the latter reflects gendered preconceptions about students' abilities in mathematics and science relative to language and humanities. In fact, I find that teachers who hold the stereotypical belief that boys are better than girls at math and science assign them higher math grades, on average, relative to girls. Similarly, teachers with stereotyped beliefs that girls are superior to boys in the humanities award them higher grades in language arts than they do to boys.

---

[2] The educational platform was created for this study with the approval of the Ministry of Education as part of a multiyear educational program. The platform, Opportunities for Everyone (Oportunidades para Todos in Spanish), is accessible at http://www.oportunidadesparatodos.pe.

[3] The average level of implicit stereotypes in this setting is considerably higher than that of middle school teachers in Italy, which is 0.09, according to Carlana (2019). In contrast, the average Implicit Association Test score of language arts teachers is 0.281, which is comparable to Carlana's 2019 findings.

Next, I estimate the impact of stereotyped teacher assessments on student long-run outcomes, such as high school graduation, college application and enrollment, and formal sector employment and earnings. My research design compares students assigned to different teachers within cohort-grade-year-school cells, controlling for a rich set of observables, including lagged test scores and students' and teachers' demographics, classroom-level controls, and school-grade-level controls. Previous studies of teacher value-added indicate that these controls are an effective means to account for the ability-driven nonrandom sorting of students to particular teachers (Chetty et al., 2014a). My research design relies on the identifying assumption that students' assignments to teachers are as good as randomly assigned when conditioning on students' demographic information and lagged test scores. I focus primarily on the effects of mathematics teachers' stereotyped grading; however, I also provide evidence regarding language arts and science teachers. I discuss the value-added of teachers in relation to high school dropout and college applications as one of the mechanisms mediating these effects. In contrast, another potential mechanism, students' assignments to sequences of stereotyped teachers, is less relevant in this setting.

My findings indicate that assignment to a mathematics teacher with stronger stereotyped assessments reflecting implicitly negative views of women's abilities in math and science relative to men, increases gender gaps in high school outcomes that persist in the labor market. Students' decisions to drop out of high school are heavily influenced by teachers' stereotypical evaluations. Girls' likelihood of graduating from high school is reduced by 1.3 percentage points (1.6% of the group mean) if they are assigned for one year to a high school math teacher who exhibits one standard deviation more severe gender stereotyped assessments regarding girls' math abilities than the average teacher. Furthermore, when female students transfer to a classroom where the teacher's stereotyped grading measure is one standard deviation higher, their likelihood of applying to college decreases by 0.7 percentage points relative to boys' (2.2 percent of the group mean). Consistent with these findings, exposure to stereotyped grading by science teachers only negatively affects high school graduation and college applications in a smaller magnitude, but it has no statistically significant consequences for employment outcomes. When girls are exposed to a language arts teacher with stronger stereotypical assessments favoring men in grading (reflecting negative implicit views of women's communicating abilities relative to men), it reduces their likelihood of graduating high school but has no statistically significant effect on their likelihood of enrolling in college.

The effects are found to be persistent and salient, as a short-lived (one-year) exposure to teachers with more salient stereotyped assessments increases the gender gap in formal-sector employment for girls aged 18–23. This research shows that female students who are assigned to a math teacher with one standard deviation stronger stereotyped assessments during one academic year are 1.3 percentage points less likely to be employed in the formal sector between the ages 18–19 (equivalent to 18 percent of the group mean). Particularly until ages 20–21, the total effect for female students is negative, and it takes young women approximately five years after graduation for the effects to level off to nearly zero. In addition, I demonstrate that female students' increased exposure to stereotyped assessments results in monthly earnings losses of USD 2.6 at ages 18–19. This effect is large enough to exacerbates the gender wage gap by approximately 32 percent, leading female students to endure cumulative losses equivalent to 0.91 percent of the mean monthly earnings during their first two years in the labor market. Finally, I investigate whether these effects are concentrated at the bottom of the earnings distribution, and I find marked and unequal losses for more disadvantaged females that last up to ages 18–19. A mechanism analysis indicates that teachers with stereotypical views about gender-math abilities manifesting in their assessments steer students toward low-paying jobs by encouraging them to internalize their own gender attitudes, resulting in persistent industry sorting three years after high school graduation.

This analysis relates to the literature on evaluators' biases and stereotypes affecting human capital investment decisions, productivity, and job performance.[4] In this body of work, started in social psychology (Schneider et al., 1979), mainly two empirical measures have been used to detect evaluators' gender stereotypes: assessment gaps constructed with observational information (Lavy, 2008, Lavy and Sand, 2018, Lavy and Megalokonomou, 2019) and the Implicit Association Test collected through surveys or laboratory experiments (Greenwald et al., 2003, 2005, 1998, Nosek et al., 2007, 2014). My focus on linking these two measures offers critical advantages compared to previous work. First, score-based measures are susceptible to capturing unobserved heterogeneity, which can be reflected in the analyzed results even after adjusting for observable drivers of assessment gaps between genders. I provide direct estimates of the robust relationship between an observed behavior, stereotyped grading, which is unexpectedly favorable to boys over girls compared to non-teacher-graded scores, and the implicit gender stereotypes favoring boys over girls in science-based fields. Second, extensive evidence indicates that teachers' grading based on classroom performance is influenced by factors other than cognitive ability or students' competencies, which poses a threat to score-based bias measures (for example, Bertrand and Pan (2013), Figlio et al. (2019), Fortin et al. (2015), Jackson (2018)). My empirical approach for measuring teachers' stereotyped assessment aims to rule out alternative explanations for gender differences in test scores, such as students' classroom behavior and ability as proxied by their past performance.

My results also add to the literature on stereotypes in education and labor contexts by measuring the effects of teacher stereotyped assessment on long-lasting human capital decisions and outcomes that have been previously unavailable. A growing literature focused on teacher-student interactions has documented that teachers' implicit biases and stereotypes based on gender disproportionately and negatively affect female students' achievement (Alan et al., 2018, Alesina et al., 2018, Burgess and Greaves, 2013, Carlana, 2019, Lavy, 2008, Lavy and Sand, 2018), choices between science- and humanities-oriented high school tracks (Carlana, 2019), or college attendance and major choice (Carrell et al., 2010, Lavy and Megalokonomou, 2019, Reuben et al., 2014). In workplace settings, evaluators' and employers' preconceived notions of workers' abilities or productivity are often taken into account when making decisions about workers' performance and job assignments (Glover et al., 2017) and job assignments (Ewens and Townsend, 2020, Goldin and Rouse, 2000, Sarsons, H., 2017).[5] I contribute to this literature by providing direct evidence that stereotypes about groups not only impair performance but also skill investments.

---

[4]From a cognitive psychology perspective, *gender biases* and *gender stereotypes* not equivalent concepts. A formal definition of stereotypes following the probability judgements approach proposed by Kahneman and Tversky (1973) is given by Bordalo et al. (2016).

This article also connects to the extensive body of research on the causes of the gender gap in the labor market. An earlier group of studies focused on the effects of childbearing (Becker et al., 2019, Berniell et al., 2021, Correll et al., 2007, Hardy and Kagy, 2018, Kleven et al., 2019), occupational sorting (Francesconi and Parey, 2018) as central causes for earnings and hiring gaps affecting female workers. Nevertheless, evidence suggests that the wage gap that cannot be accounted for by workers' skills or education levels persists (for instance, see Bertrand et al. (2010, 2005), Blau and Kahn (2017), Goldin (2014)). Recent research has shifted its focus to the psychological and behavioral causes of the persistence of gender gaps. Among these, the existence of differential preferences for job attributes such as schedule flexibility and stability (Fleche et al., 2018, Wiswall and Zafar, 2018), commuting time to the workplace (Le Barbanchon et al., 2020) or competitiveness preferences (Flory et al., 2015, Gneezy et al., 2003, Niederle and Vesterlund, 2007) have proved to be important causes. Female workers' behavioral traits, such as the propensity to bargain (Biasi and Sarsons, 2020), to ask for higher wages (Rousille, 2021), confidence (Risse et al., 2018), or the subjective description of performance in tasks (Exley and Kessler, 2022), have opened new routes to explain the remaining unequal access to wages and hiring opportunities. I contribute to this literature by documenting that high school students internalize their teachers' stereotyped assessments regarding numeric and communication abilities. This is a key mechanism by which these stereotyped views can have lasting effects on their employment prospects and earning potential. Female workers who are exposed to teachers with more robust stereotypical views about women's numerical abilities as students develop biases against women in mathematics and science, as measured by students' IAT scores. This internalization may affect their investments in academic careers or cause them to choose not to work in math-related industries, which is reflected in the systematic industry sorting patterns of workers exposed to instructors with more stereotyped grading. Notably, teachers' biases are a previously undocumented source of gender disparities in labor markets, suggesting that the origins of such gender gaps were, in part, formed prior to workers' entry into labor markets.

Last, the paper builds on the empirical literature on discrimination in labor markets and educational settings. Methods to retrieve precise individual-level estimates of discriminatory practices and the distribution of those estimates have a recent precedent in this body of work (Kline and Walters, 2021, Kline et al., 2021). This paper contributes to refining the methods we use to measure gender stereotypes and bias in education and has potential uses in other domains. I rely on an Empirical Bayes approach used in the statistics literature (Efron, 2012, 2016, Efron and Morris, 1972) to measure heterogeneity in discriminatory behaviors across teachers using assessment-based jointly with psychological measures of gender biases.

The remainder of this paper proceeds as follows. Section 2 provides a framework that conceptualizes teachers' stereotyped assessments and mechanisms, and it introduces a set of economic and psychological studies that motivate the empirical exercises. Section 3 introduces the empirical setting and describes the data. Section 4 presents the estimation strategy for measuring teachers' stereotyped grading, characterizes the distribution of stereotyped teacher assessments among high school teachers, and provides validation using IAT scores. Section 5 describes the methodology for measuring the long-term consequences of teachers' stereotypical assessments. Section 6 investigates the effects on academic progress and college outcomes, and Section 7 describes the effects on labor outcomes. Section 8 analyzes the consequences of teachers' stereotyped assessments in subjects different from mathematics. Section 9 concludes.

---

[5] Glover et al. (2017) finds that minority workers under the supervision of managers with higher implicit stereotypes against them, as measured by the IAT, had lower performance metrics . Sarsons, H. (2017) documents an asymmetric penalization of surgeons' past performance measured by patient death that depends on physicians gender when determining subsequent job assignments (that is, surgical referrals).

## 2 A conceptual framework for human capital investment with stereotyped assessment

The framework in this section illustrates the theoretical mechanisms by which gender-prejudiced attitudes, stereotypes, or biases can affect human capital investment decisions. This discussion theorizes gender stereotypes in assessment and their persistent influence on students' outcomes; therefore, it motivates the empirical analysis of the following sections. I draw from a growing body of studies in economics, educational and social psychology that assess how gender or racial stereotypes and discriminatory behaviors affect individual decisions and how they have tangible ramifications.

The most explored mechanism of teachers' gender stereotypes is their direct effect on student achievement, which is the foundation for subsequent student learning. An early literature on students' achievement gaps in mathematics is a crucial starting point.[6] Recent investigations show how gender biases affect such achievement gaps (Alesina et al., 2018, Carlana, 2019, Lavy, 2008, Lavy and Sand, 2018). In particular, Carlana's (2019) study highlights the effects of teachers' implicit stereotypes, which potentially extend to postsecondary fields of study for female students, as they prevent them from choosing science-oriented tracks before high school. I turn to the literature that evaluates the effect of racial or ethnic stereotypes on student performance. For instance, van den Bergh et al. (2010) find that teachers' ethnic stereotypes, as measured by the IAT and self-reported measures, affect the academic achievement of minority groups. Similarly, using measures of racial stereotypes in assessment, Botelho et al. (2015) document a grading gap between minority and nonminority students that is not accounted for by observable proficiency or student behavior.

The associated change in expectations toward the affected student groups is the cornerstone of the educational research investigating the mechanisms of teachers' stereotypes. The relevance of teachers' expectations has been established by a longstanding body of work in education (Babad et al., 1982, Rosenthal and Jacobson, 1968), and the work continues in recent studies that focus on the asymmetric effects of positive versus negative expectations, the accumulation of expectations of multiple teachers over time, and related questions (see a recent review in Jussim and Harber (2005)).[7] Some of this work analyzes the influence of teachers' expectations based on group membership (for instance, ethnicity; see Rubie-Davies et al. (2006), van den Bergh et al. (2010)) or expectations based not on individual characteristics but on characteristics of aggregates such as the class (Rubie-Davies, 2010). Group membership-based standards shape teachers' group-specific performance expectations. Such standards align with economics research that either theorizes (Coate and Loury, 1993) or empirically documents (Burgess and Greaves, 2013) that teachers' gender or ethnicity-based expectations may become self-fulfilling.

---

[6] For instance, literature reviews on the gender achievement gap can be found in Bertrand (2011), Fryer and Levitt (2010), Hyde and Mertz (2009).

[7] Other relevant reviews on the literature are in Harris and Rosenthal (1985), Hoge and Coladarci (1989).

If not through expectations, through which alternative channels can teachers' biases impair the academic progress and adult outcomes of students? Teachers' disproportionate allocation of educational resources has been found to affect students' academic achievement (Alan et al., 2018, Carrell et al., 2010). Recent work by Alan et al. (2021) sheds light on how teachers' ethnicity-based stereotypes affect the distribution of seats in the classroom, leaving some students to receive less instructor attention and fewer opportunities for peer connections. Next among teachers' stereotype channels is the association of stereotypes with educator quality, as shown in the value-added literature (Chetty et al., 2014a, Jacob and Lefgren, 2008, Kane and Staiger, 2008, Rothstein, 2010) and in the context of biased grading Lavy and Megalokonomou (2019).

Behavioral and psychological traits are another set of mechanisms being investigated. According to studies in educational psychology, teachers' stereotypes have been found to affect students' self-assessment or the formation of their academic self-concept (Ertl et al., 2017). Exley and Kessler's (2022) found that female workers subjectively assess their performance in so-called male-typed math and science tasks as supbar relative to the average self-assessment of male workers. Another recent body of research in social psychology investigates internalized stereotypes, a psychological construct by which minorities or members of groups subject to negative stereotypes, beliefs, prejudices, and other forms of discrimination accept the negative stereotypes and beliefs about themselves and their group and, as a result, exhibit this behavior (David et al., 2019, Jones, C. P., 2000).[8] I follow a growing social- and clinical-psychology literature that investigates the relationship between racial discrimination and psychological distress (see Molina and James (2016), Roberts and Rizzo (2021), Willis et al. (2021)).

In this study, I hypothesize that educators and students can internalize gender stereotypes (for instance, female teachers internalize gender-science stereotypes against their own numeric abilities and act accordingly). Consequently, teachers who exhibit gender stereotypes directly or have internalized them can perpetuate their pervasive effects on their students, causing them to internalize them. In other words, students may be led to change their self-perception of their educational adequacy or of their ability to complete mathematical learning tasks or scientific endeavors. This influences subsequent high school dropout decisions and manifests itself in other adult outcomes.

Gathering the main channels introduced in this discussion, I formalize how teachers' gender stereotypes shape students' decisions to acquire qualifications (for example, a high school diploma or a college degree) in Online Appendix B. I develop a dynamic model of human cap-ital investment from the students' perspective, as a discrete choice regarding whether or not to acquire a qualification in an environment of incomplete information about their academic potential.[9]

---

[8]The work of Jones, C. P. (2000) started this literature. It develops a conceptual framework that postulates that racial discrimination works on three levels: institutionalized (that is, unequal access to public services and opportunities due to race), personally mediated (for example, prejudices, stereotypes, racial discrimination), and internalized racism, defined as the "acceptance by members of the stigmatized races of negative messages about their own abilities and intrinsic worth."

In this model, students from two groups that are fully distinguishable to teachers (for example, boys and girls) develop self-notions about their academic potential based on their teachers' and a central planner's assessments. I allow teachers' assessments to reflect their stereotypes about the distribution of students' abilities across groups. I assume that students are rational decision-makers who maximize the utility derived from acquiring qualifications after accounting for their costs. Although qualification expenses vary across students, they are determined identically in both groups. The central result of the model is the implications of teachers' gender stereotypes for the equilibrium self-assessment of their students. In particular, I derive conditions for an equilibrium of self-fulfilling negative gender stereotypes against either of the two groups, demonstrating that students against whom teachers hold negative gender stereotypes have suboptimal self-assessments, which, over time, cause them to underinvest in their academic trajectories compared to students from the group regarding which teachers have a positive view of their abilities. In the following sections, I obtain empirical estimates of teachers' gender stereotypes in assessment and their long-term consequences, allowing me to empirically assess the prevalence of a self-fulfilling equilibrium as predicted in Online Appendix B.

## 3 Background and data

### 3.1 Public school system in Peru and gender equality

The Peruvian public school system provides free education to 5.9 million students, representing 74.5 percent of the total student population in 2019. The public school sector is the largest employer of teachers in the country, employing 69.7 percent of teachers full-time or part-time. The other main type of education provider is private schools, which do not operate within a voucher or charter system. Schooling is compulsory between ages 5 and 16, and it is divided into three levels: preschool (ages 3 to 5), primary (ages 6 to 11), and secondary (ages 12 to 16) (UNESCO, 2016). I focus on the secondary level, in which 1.9 million students were enrolled from grades 7 to 11 in 2019.

Public school enrollment is available to all school-age children; however, those who enroll tend to come from underprivileged backgrounds. For instance, the 2018 National Census Evaluation questionnaire of eighth-grade students indicates that 43 and 27 percent of parents answered that their highest level of education was less than high school completion and high school completion, respectively (see Appendix Figure C.1). Moreover, the questionnaire indicates that the typical eighth-grade public school student in 72 percent of districts nationwide has very low socioeconomic status (see Appendix Figure C.2).

Even though access to education increased steadily between 2010 and 2019, with attendance increasing at an average annual rate of 1.1 percent, school completion still represents a challenge, especially among female students outside of metropolitan areas in the east or northwest (see Appendix Figure C.3). Recent reforms have aimed to increase the quality of education provision and provide equal opportunity for female and male students.[10] In 2016 the Ministry of Education introduced a new national curriculum that takes an inclusive approach to diversity, with a guiding principle of gender equity and inclusion (Minedu, 2016). However, it has not materialized in high school or other pedagogical content. Implementation is widely left to the discretion of regional and local units equivalent to school districts.

---

[9]The model builds on Coate and Loury's (1993) model of self-reinforcing prior beliefs.

Gender-equality attitudes in Peru regarding, for instance, access to education or employment have not progressed to match those seen in developed countries.[11] As I show in the following sections, this trend is pervasive in the education setting.

## 3.2 Data and samples

In this section, I describe the two main samples I use for the empirical analysis and I describe their data sources. The first sample is used to estimate a measure of stereotyped assessment for each teacher. The second sample contains student information, which I use for evaluating the effects of stereotypical grading practices on a set of adulthood outcomes.

### 3.2.1 Student sample: Estimating teachers' stereotyped assessments

The measure of stereotypical teacher assessments of girls measures which teachers are more likely than others to assign examination grades that, on average, are unfavorable toward female students relative to male students, using blindly graded test scores as the benchmark. The parameter is a double difference, with differences between two types of examinations and between gender-defined student groups. To retrieve teacher-level measures, I construct a sample of students whose end-of-year classroom scores and standardized test scores are matched with their respective teachers. In the empirical context of Peru, high school teachers tend to specialize in a single subject but teach a variety of grades and school years. I use this turnover to calculate a measure of stereotypical grading practices for eight grade levels from 2015 to 2019. I then assign these measures to other classrooms that the teacher instructed in the same subject and time period (see Online Appendix A for the detailed procedure). An important assumption is that the gender stereotyped grading practices exhibited by a teacher teaching eighth-graders is not different from the teacher's stereotyped grading when they teach other grades.

The teacher in charge of instructing a subject is responsible for assigning a consolidated mark at the end of the school year. The grades assigned by teachers do not reflect the performance of students on a standardized test with a single format. Instead, they reflect the proficiency of students at the end of the school year in a set of competencies for a set of learning goals, as measured by multiple examinations using a rubric established by the National Curricula.

---

[10] In addition to gender, geographic location and race are factors contributing to unequal access to education services. Among the reforms to improve achievement gaps and overall education quality, there have been reforms to introduce an evaluation-based selection process for teachers (Bertoni et al., 2022), establishing payment and incentive schemes for staffing schools in rural areas (Bobba et al., 2021), establishing auditing information systems preventing and reporting bullying (Minedu, 2022), and improving school equipment and infrastructure (Paxson, 2002).

[11] The WVS-7 results for Peru suggest a prominent level of gender-inequality attitudes regarding equal access to education for women. 12.9 percent of respondents agree with the statement *'University is more important for a boy than for a girl'* in Peru, compared to 9.8 percent for the US. The score for gender equality in education is on a 0-to-1 scale, where higher values indicate more favorable attitudes toward equality.

In contrast, standard scores reflect the performance of students on a single assessment. The Ministry of Education annually administers a nationwide standardized test to eighth-grade students in mathematics, language arts, and science in three exams at the end of the school year.[12] The format and content of the standardized and classroom evaluations are described in Online Appendix F. Table 1 shows descriptive statistics of the sample of eighth-grade pupils whose scores are used to compute a measure of teacher-level assessment bias. Students are, on average, 13 years old. More than half of the sampled students speak Spanish as their primary language, 20 percent were born in Lima, and the schools that 86 percent of them attend are in urban areas.[13] A large number of these students come from low-income families, and around 24 percent receive a conditional cash transfer.

### 3.2.2 Student sample: Long-term outcomes

**Data sources.** This analysis uses several data sources to estimate the effects of teachers' stereotyped assessments on a sample of 1.7 million public high school students graduating between 2015 and 2019. The data set from which I draw that sample tracks students' exposure to stereotyped grading practices mathematics, language arts, and science teachers between grades 7 and 11. It tracks their high school–completion, college-attendance, and labor market outcomes between ages 18 to 23, depending on the graduation cohort. Panel A of Appendix Table C.1 illustrates the grades for which I can observe gender stereotypical grading and the period over which I can measure outcomes for each student cohort.

The main data set I use for constructing this sample is the universe of public school students' administrative records, including students' classroom assignments, evaluation score records, enrollment records, demographics, high school–completion records, and college records provided by the Ministry of Education. I also use employer and employee information provided by the Ministry of Labor to determine whether students obtain employment on the formal labor market after graduation, as well as to calculate earnings and working hours. In addition, I use comprehensive information on teachers' classroom assignments, which includes their demographic information, course subject, and performance score, from a national teacher evaluation for hiring public school instructors in 2015, 2017, 2018, and 2019[14]

**Sample construction.** High school students are assigned one homeroom teacher per classroom and have subject-specific teachers during the entirety of the school year. Therefore, students' and teachers' classroom-assignment records are linked at the classroom-subject level for each grade using classroom identifiers. The purpose is to calculate the exposure of students to stereotypical assessments in each subject and grade. The successful student-teacher matches constitute my base sample. Next, I join with the information on estimated teachers' stereotyped grading using teacher identifiers.[15] Finally, after applying restrictions to preserve valid matches in the sample, I link this data set with the students' long-run outcomes using unique student-level identifiers. This constitutes the full sample that I use for my analysis below (see Online Appendix A for details).

---

[12] The National Standardized Evaluations of high school students administered in 2015, 2016, 2018, and 2019 were scheduled in November, and the school year finished in the second week of December at the latest. I do not use the information from 2017, as the standardized evaluation was canceled for that year because of a natural disaster in the country.

[13] The primary-language statistics for this sample of eighth-grade students differ from the primary-sample descriptive statistics in Table 2 because of missing values in the administrative data set regarding the takers of national standardized tests.

[14] The National Teachers Evaluation was established in 2014 to assess the competencies of teachers aiming to attain permanent positions and access a regulated payment scale and additional supplementary wage bonifications, retirement plans, and promotion schemes.

**Outcomes.** The primary outcome variable is high school completion, which I measure over different time frames that allow me to keep track of the academic progress of students. Similarly to Gray-Lobe et al. (2021), I construct a projected graduation year, defined as the year the student is expected to graduate in a normal academic progression after five years of high school. To calculate this variable, I take as a reference year the school year when the student is enrolled in grade 7, and I compute the year in which the student will graduate if they advance to grade 11 without retention. For instance, a student in cohort I who is in grade 7 in 2015 will have 2019 as their projected graduation year. Using the *projected* variable, I define *on-time* high school graduation to mean that students finish in the projected graduation year and *ever* graduating from high school to mean students finish at some point, whether on time or late.[16]

The second set of outcomes consists of college application, admission, and enrollment. As before, these are defined using the on-time and ever concepts, which enables me to evaluate the impact of the gender stereotypical assessments of teachers not only on college attendance but also on deviations from the standard progression. The final set of outcomes captures the experience of public high school graduates venturing into the labor market with a high school diploma. In particular, the outcomes measure whether the student is employed in the formal sector and what their annualized earnings and annualized work hours are. The time frame for tracking the evolution of these outcomes is between ages 18 to 23 after graduation. Given that some students who do not manage to graduate from high school still participate in the labor market, I consider the projected high school graduation year as the common threshold for all students.

I estimate annual employment in the formal sector, earnings, and work hours using employment records that report these outcomes monthly—a higher frequency than the quarterly or annual measures available in many administrative data sets (for instance, see Card et al. (2013), Sorkin (2018)). The granular data allows me to determine whether a worker is employed in the formal sector despite the precarious nature of work in emerging economies and the high crossover between the formal and informal sectors (OECD, 2015). I consider a worker to be employed in the formal sector if they hold a job contract with their primary employer with positive earnings in the formal sector for at least one month with their primary employer between 2015 and 2020.[17] A worker's *primary employer* is defined as the employer that pays the largest share of the worker's earnings in a quarter (Abowd et al., 2003, Lachowska et al., 2020, Sorkin, 2018). This definition is convenient for considering the scenario of multiple employers for a single worker and for retrieving annualized earnings and work hours. Online Appendix A explains the construction of the labor-outcome variables in detail.

---

[15]Section 3.2.1 describes the sample used to construct the stereotyped grading measure, and Section 4 describes the empirical strategy for using teacher-assigned scores for constructing a measure of stereotyped teacher assessments.

[16]Dropout status can change, as students may re-enter school an indefinite number of years after leaving it. My definition of high school completion considers this re-entry, but it does so over different time horizons for each cohort because data availability is limited. For example, for students who drop out in 2015, I can see whether they return during 2016–19, while if someone dropped out in 2018, I can only see whether they return during 2019. In addition, as seen in Appendix Table C.1, the data structure is an imbalanced panel so I observe each cohort of students in a different number of grades over high school. For instance, I observe cohort I only over grades 10 and 11, while cohort V is observed over grades 7 to 11. Nevertheless, the starting school year for calculating the *projected* graduation year will not be the seventh grade for all cohorts for which I use the student's enrollment records.

**Descriptive statistics.** I study outcomes of five cohorts of students at two points in time: during their high school years and after graduation. Table 2 presents summary statistics of demographic and education characteristics during high school for the (stacked) base and full samples of students in grades 8 to 11. Information is available for these student cohorts starting in seventh grade, but the empirical strategy requires lagged test scores, so the analysis sample starts in eighth grade. The table allows a comparison between the characteristics of students in the base sample, which consists of successful student-teacher matches, and the characteristics of all students with a stereotyped grading measure. Students in the complete sample are more likely to speak an indigenous language (Quechua), to have been retained in high school, and to have been born outside of Lima. Over eighty percent of public school students speak Spanish, and approximately two out of ten students were born and educated in Lima.

Once these students leave the schooling system, I observe their workers' records for 2015–20. Each cohort in the sample is observed in the employer-employee records for a different time window based on their projected graduation year. Panel B of Appendix Table C.1 displays the cohort-year cells for which I can observe labor market outcomes. Table 3 shows descriptive statistics for the sample of public school students matched to labor records in Columns (4) and (5) and a comparable benchmark group of workers between 18 and 25 years old from the entire population in Columns (1) to (3). The first panel in the table shows average earnings and hours computed using the monthly information for 2015–20, and Panel B only reports workers' characteristics for the most recent employment relation with an *dominant annual employer* as defined in Online Appendix Section A.2.1. The full regression sample includes 485,000 workers with available records and a projected graduation year between 2015 and 2019. This group represents roughly 20 percent of the base sample. Female workers in the regression sample have around 15 percent lower wages than male workers, and a similar gap is observed in the population. The rate of workers reporting that they finished high school is similar across gender groups, with a higher rate of men reporting having attained some college or technical education.

## 3.3 Gender-attitudes survey and the Implicit Association Test

### 3.3.1 Sampling and survey design

My survey aimed to quantify teachers' gender attitudes and implicit stereotypes as measured by the IAT, as well as evaluate the relationship between IAT scores and the stereotypical assessment measure. Figure 2 shows the data-collection timeline, which started in September 2021 and ended in September 2022. The sampling frame comprised high school teachers appearing in

---

[17]As detailed in Online Appendix A.2.1, I also consider the job spells for two and three consecutive months.

Ministry of Education administrative records that included contact information and school information according to the School Census 2019. During most of the data collection period, school activities were not in person because of the COVID-19 pandemic. Therefore, contact information (phone numbers) needed to be part of the sampling frame, as survey participants had to be contacted via text messages and phone calls. Approximately 230,000 teachers had available school and contact information matching the sampling criteria.[18] I conducted stratified sampling by region and selected 13,654 teachers to receive a text message inviting them to participate on a web-based platform hosting the survey. Each sampled teacher received a text message in two waves, one in October during the 2021–22 school year and another in January during the 2021–22 school year. Figure 1 illustrates the completion rate of teachers and the data coverage nationwide.

The survey was available on a web platform I designed and deployed for the Ministry of Education as part of its remote-learning program targeting high school students. The platform is part of a multi-year effort to study gender gaps and inequality sources affecting young students' trajectories.[19] During the data-collection period, 8,687 teachers registered to participate and 2,462 teachers completed the entire survey. Teachers who visited the website had to give informed consent in order to register and participate in the activities. No reference to gender stereotyped assessments was made in the consent form or on the web platform. The goal communicated to teachers was that the project aimed to gather information on teachers' practices in order to promote equal opportunity for all students at public schools. The survey was divided into two activities: a teachers' questionnaire and an association test.

My final survey sample includes 1,102 mathematics and 950 language arts teachers (N= 2,052) who completed both activities and were successfully matched to the teacher-student matched sample that contains a measure of stereotyped teacher assessments. Teachers from other subjects who completed the survey and the IAT were excluded from the analysis sample because information on stereotyped assessments was unavailable.

### 3.3.2 Teachers' questionnaire and Implicit Association Test

The IAT is a psychology test developed to measure implicit stereotypes (Greenwald et al., 2003, 1998, Nosek et al., 2007, 2014). Unlike self-reported gender-related attitudes, IAT responses are less influenced by cognitive processing, as they represent an indirect measure of stereotypes: the speed of categorizing science and gender terms (De Houwer and Moors, 2007, De Houwer et al., 2009). Therefore, the IAT is a more reliable assessment of implicit attitudes and overcomes social desirability bias (Egloff and Schmukle, 2002, Steffens, 2004).[20]

---

[18]The number of teachers for which the Ministry of Education had information matching the sampling criteria was taken from personnel records by government officials with clearance to access teachers' contact information.

[19]The platform's name, Opportunities for Everyone (*Oportunidades para Todos* in Spanish), does not make explicit references to gender stereotypes. However, it appeals to teachers and students to report information for the ministry to develop learning tools and policies that will promote learning in diversity and inclusion for all students.

[20]Socially desirable responding is a type of response bias in which participants answer questions aligning to cultural norms that let them be accepted or appraised positively by others. Such responding occurs despite their true latent attitudes (Fisher, 1993, Paulhus, 1991, Podsakoff et al., 2003). It has been documented to affect job-application behavior (Stöber, 2001) and political poll responses about race-related policies (Berinsky, 1999, Krysan, 1998).

A growing literature finds evidence correlating IAT measures with minority job applications (Rooth, 2010), teachers' high school track recommendations (Carlana, 2019), and minority workers' job performance (Glover et al., 2017). An active area of study in psychology concerns the capacity of IAT-measured stereotypes to predict observed or self-reported behavior (Buttrick et al., 2020, De Houwer and Moors, 2007, De Houwer et al., 2009). Nevertheless, similarly to recent findings in economics, studies in psychology document a relationship between IAT-measured bias and various observed behaviors, such as health care services by practitioners (FitzGerald and Hurst, 2017) and harmful intergroup behaviors (Rudman and Ashmore, 2007).

For this study, I administer two IATs in random order to participants: the gender-science IAT and the gender-career IAT (see details on the IAT in Online Appendix Section D.1). In the former, teachers are asked to associate science and humanities with gender groups, while in the latter, they associate career and family-related terms with gender groups. More specifically, the gender-science IAT measures teachers' strength of association between science and humanities concepts, on the one hand, and group terms referring to girls and boys, on the other hand. The test instructs participants to make associations as fast as they can between words from each group (that is, sciences and humanities) and gender terms (that is, female and male), measuring the association speed in milliseconds. Quicker associations between, for example, humanities and girls or sciences and boys demonstrate the participants' stronger latent connection between these two, reflecting implicit stereotypes that exclude women from science disciplines.

The IAT score is a standardized measure with a possible range between -2 and 2, where higher values indicate a stronger implicit preference for boys in science careers and girls in humanities careers, relative to boys in humanities and girls in science. I follow the improved scoring procedure proposed by Lane et al. (2007) and Greenwald et al. (2003), which discards teachers' overly slow or fast responses to prevent contrived or fatigued random associations from being included in the analysis. Throughout the following analysis, interpretation of the IAT's gender-stereotypes severity is based on the following discretizing IAT categories: *preference for girls*, *little to no stereotypes*, *slight stereotypes*, *moderate to severe stereotypes*, and *strong stereotypes*.[21]

Additionally, the teachers' questionnaire collects complementary information to that obtained in the IAT, particularly self-reported gender attitudes and information on how teachers exhibit implicit gender stereotypes in the classroom. Online Appendix Section D.2 has further information on the teachers' questionnaire.

### 3.3.3 Descriptive statistics

Table 4 presents information on the surveyed teachers who have been matched with a measure of stereotyped grading and therefore constitute my analysis sample. Several participant characteristics bear emphasis. Gender stereotypes are pervasive in the sample of surveyed teachers, as 43.4 percent of mathematics teachers have a moderate to severe or strong stereotypes against girls in science, while the same holds for 45.8 percent of language arts teachers. Regarding career stereotypes, 40 and 38.4 percent of mathematics and language arts teachers, respectively, exhibit such levels of stereotypes against girls in professional career trajectories. More mathematics teachers are men in the analysis sample, while the remaining demographic characteristics are roughly the same across mathematics and language arts teachers and across school locations. Mathematics and language arts teachers work around 28.1 hours per week, and the class size is, on average, 25 students across both subjects. Forty percent of teachers have a tenure contract; on average, teachers in public schools have 13 years of experience. Last, consistent with the level, 16–18 percent of teachers have witnessed or experienced discrimination firsthand perpetrated by other colleagues or school principals.

---

[21]The break points for determining the categories are $-.15, .15, .35, .65$ as proposed by Greenwald et al. (2003, 2005).

## 4 Measuring teachers' stereotyped assessments

In this section, I construct a measure of systematic gender differences in teachers' grading that conveys gender-math stereotypes favoring boys in math-science and women's talent in the humanities using student-teacher matched data for eighth grade students. I use the IAT to examine the hypothesis that teachers have gender-specific prior beliefs about the ability distribution of their students in math-science, and communicating abilities in this setting, and to what extent these stereotypes predict gender differences in grading across teacher- and blindly-graded examinations.[22] The result is a relative metric that allows me to compare the degree of stereotypical assessment across teachers.

### 4.1 Estimation strategy

I define systematic gender differences in teacher $j$'s assessment as the disparities between male and female students' mean gaps between teacher-assigned scores and blindly graded test scores. Higher values of this measure indicate teachers are, on average, grading male students more favorably than female students, while lower values indicate the opposite. The set $\mathcal{N} = \{1, \ldots, N\}$ denotes a student population indexed by $i$; teachers indexed by $j$ are part of a population denoted $\mathcal{J} = \{1, \ldots, J\}$. Students are assigned scores by their teachers and a central planner (for example, the school-district authority or the agency in charge of standardized examinations). $S^T_{ij}$ denotes the examination assessment of student $i$ assigned by teacher $j$. In addition, students receive another score, denoted $S^B_{ij}$, as measured by a standardized national examination.

I define a system of estimating equations, with parameters of interest $(\alpha_{1,j}, \alpha_{2,j}, \beta_{1,j}, \beta_{2,j})$:

$$S^B_{ij} = \alpha_{1,j} + \alpha_{2,j} M_i + W'_{ij} \alpha_3 + \eta_{ij} \tag{1}$$

$$S^T_{ij} = \beta_{1,j} + \beta_{2,j} M_i + W'_{ij} \beta_3 + \epsilon_{ij} \tag{2}$$

The specification includes teacher-specific intercepts $(\alpha_{1,j}, \beta_{1,j})$ and gender effects on score assignment $(\alpha_{2,j}, \beta_{2,j})$. Under the identifying assumptions, the difference between parameters $\beta_{2,j}$ and $\alpha_{2,j}$ averaged by teacher measures the gender stereotypes that enter into their competence assessments. Covariates $W_{ij}$ include student characteristics (for example, student age in months and place-of-birth indicators), quadratic polynomials of lagged scores in mathematics and language arts, and quadratic polynomials of lagged physical education scores as proxies for observed classroom behavior as in Botelho et al. (2015). I estimate Equations (1) and (2) separately via a fixed-effects approach using the teacher-student matched sample containing classroom grades and standardized assessments in the respective subjects. I report robust standard errors clustered at the student level. In Section 4.2, I use these parameters to construct a measure of teacher-level systematic gender differences in assessment.

---

[22] Appendix B introduces a framework that formalizes the influence of teachers' gender stereotypes on human capital investments.

Omitted variable bias is a challenge when measuring gender differences in grading to reflect gender stereotypes. Boys could be assessed systematically higher or lower than girls on teacher-graded tests, not because teachers are biased but because of unobservables correlated with gender. Some possible factors driving better performance by either group include teachers consistently prioritizing content favoring one group or having a teaching style that does so; teacher scoring reflects student discipline in the classroom (Botelho et al., 2015); test-taking skills differ depending on exam format (Ben-Shakhar and Sinai, 1991, DeMars, 1998, 2000, Gallagher and De Lisi, 1994, Klein et al., 1997, Liu and Wilson, 2009); and examination conditions favor one gender over another (Gneezy et al., 2003, Niederle and Vesterlund, 2010). I address this concern by controlling for comprehensive student and teacher characteristics to adjust the female-male score gaps.

## 4.2 Teacher-level gender differences in assessments

The teacher-level differences between teacher-assigned and centrally assigned scores attributable to the disclosure of students' gender in non blind examinations are characterized as follows:

$$\theta_j := \beta_{2,j} - \alpha_{2,j} \qquad (3)$$

I interpret $\theta_j$ as the adjusted discrepancies in gender gaps from teacher-graded versus centrally graded assessments measured in standard deviation units. When this underlying parameter of interest takes higher values for teacher $j_1$ relative to teacher $j_2$, it indicates that $j_1$ is more likely to assign higher scores to boys than to girls compared to the gender gap on blindly graded tests because of the teacher's ability stereotypes against girls. Thus, this is a relative measure, allowing for comparisons of gender differences in assessment across teachers.

There are two restrictions that allow me to identify the parameter of interest in Equation (3) and to interpret it as an informative measure of stereotyped assessments for comparisons across teachers. First, I assume there are no unobserved teacher-by-gender factors in student skills that differentially affect the gender gap on teacher assessments compared to blindly graded scores. Additionally, I assume that if there is any skill differential between boys and girls, it does not vary by teacher across their classrooms. These are essentially the assumptions needed to identify parameters in a difference-in-differences research design in which I compare boys versus girls and teacher-assigned versus blindly graded examinations. These assumptions do not rely on a cardinal interpretation of $\theta_j$; that is, the location of $\theta_j$'s distribution need not be interpreted as a causal impact of gender stereotypes. I discuss potential threats to the identification of $\theta_j$ in the following section.

### 4.2.1 Heterogeneity in teacher-level gender differences in assessments

The teacher-level estimates $\hat{\theta}_1, \ldots, \hat{\theta}_J$ and associated standard errors denoted $s_1, \ldots, s_J$ are obtained by estimating Equations (1) and (2). The scores used for estimating $\hat{\theta}_j$ have been standardized by year and subject. Table 5 summarizes the distribution of estimates of teacher-level gender differences in assessment for mathematics, language arts, and science teachers. The first row reports the mean gender differences in grading by subject without any adjustment, $J^{-1} \sum_j \hat{\theta}_j = \hat{\mu}$, for teacher $j$ of a given subject.[23] The degree of stereotyped assessments varies considerably across teachers. Appendix Figure C.4 displays the distribution of $\theta_j$ for all three subjects; it indicates that in all subjects, some teachers are systematically assigning higher scores to boys, while others are systematically awarding higher grades to girls.

Next, I analyze the heterogeneity of gender differences in teacher assessment. Due to sampling error, the estimates $\hat{\theta}_j$ are expected to be more variable than the underlying parameters $\theta_j$. As a result, the variance of the estimated distribution of $\theta_j$ is likely upwardly biased. This sampling variation can have several sources. For instance, for teachers that have limited experience or small classrooms, the estimated $\hat{\theta}_j$ are very imprecise because the number of observations is limited (see, for example, Kane and Staiger (2002)). Therefore, a bias corrected estimate of the underlying $\theta_j$'s variance is $\hat{\sigma}^2_U = J^{-1} \sum_j [(\hat{\theta}_j - \hat{\mu})^2 - s_j^2]$.[24] Additionally, I compute a student-weighted version of the bias-corrected variance to reduce the noise estimated in $\hat{\theta}_j$ that is related to a very small class size. That variance is defined as $\hat{\sigma}^2_W = J^{-1} \sum_j \left[w_j(\hat{\theta}_j - \hat{\mu})^2\right] - \sum_j w_j s_j^2$, where student weights are $w_j = \mathcal{N}(j)/\mathcal{N}$ and $\mathcal{N}(j)$ is defined as above for teacher $j$ for a given subject and $\mathcal{N} = \sum_j \mathcal{N}(j)$ for that subject.

The second and third rows of Table 5 indicate that mathematics teachers exhibit the lowest dispersion in gender differences in assessment. The bias-corrected weighted standard deviation of systematic gender differences in grading, to the disadvantage of girls, ranges from 0.06 to 0.09 test-score standard deviations for mathematics teachers. Consistent with these results, the dispersion of the systematic differences in grading boys and girls by mathematics teachers is also smaller than that of language arts teachers in Lavy and Sand's (2018) findings.[25] In contrast, in my analysis sample, language arts teachers' dispersion of gender grading differences against girls is lower than that of their science colleagues: moving upward one standard deviation in the distribution of language arts and science teacher gender grading gaps increases the score gap of boys versus girls by 0.11 and 0.16 test-score standard deviations, respectively. The lower dispersion of gender assessment differences in mathematics could be because instructors have limited discretion to grant partial credit on numerical exercises, or it could be that math

---

[23]If I interpret $\hat{\theta}_j$ as a measure of *absolute* stereotypical grading instead of a relative one, the negative mean would imply that, on average, teachers hold stereotypical grading practices against boys. However, this *absolute* interpretation is not credible, because the classroom and standardized examinations could be testing slightly different skills. For instance, boys could be slightly worse on average at those skills because of differences in assessments or unobserved skill gaps across gender, as argued in the following paragraphs. This would shift the mean gender discrepancies and, therefore, shift $\vartheta$. Nevertheless, the *relative* interpretation remains valid as long as that skill differential does not vary by teacher (as required by the second identifying assumption). Moreover, under the triple-difference interpretation of $\hat{\theta}_j$, the location of the $\vartheta$s does not play a relevant role in my analysis.

[24]Similar bias-corrected-variance procedures where estimated standard errors, $s_j$, are subtracted to depurate the observed variance can be found in Aaronson et al. (2007), Abdulkadiroğlu et al. (2020), Angrist et al. (2017), Kline et al. (2021).

[25]Strictly, the English and Hebrew teachers reported in Lavy (2008) are referred to here as language arts teachers.

students can more easily verify whether their answers are correct.

**Discussion of identifying assumptions.** The main potential threat to identification stems from the fact that teacher-assigned and centrally assigned scores could reflect the evaluation of slightly different skills among boys and girls. The possibility of different assessments across examinations is largely rejected based on the analysis contained in Online Appendix F, as I compare the contents evaluated in both types of examinations and I analyze the distribution of contents across questions in the standardized tests and typical classroom examinations in mathematics and language arts.[26] My content analysis shows there are no major differences in the distribution of learning competencies evaluated across the examinations taken by students at the end of eighth grade.

A related threat to identification is that, in some classes, boys (compared to girls) could be more likely to have more of the skills that are required for the teacher-assigned tasks and not the centrally graded tasks. To understand this concern, consider a scenario in which observed and unobserved skill gaps across genders might exist. For instance, cultural norms could motivate boys in some classrooms of a given teacher to allocate their effort to improving their skill at rapid problem solving to earn them higher marks in the types of questions that are common in classroom examinations. In fact, a long-standing literature in educational psychology explores how different test formats present different challenges across genders and racial groups. For instance, Liu and Wilson (2009) investigate the existence of gender differences in performance by type of exam question. The authors conclude that male students have an advantage over female students on complex multiple-choice items and other question types in which the evaluated content departs from textbook context or requires a new use of standard methods learned in class. This aligns with the research of Gallagher and De Lisi (1994), who documents that girls outperform in questions related to textbook content. Moreover, Klein et al. (1997) finds evidence that girls have better outcomes than boys in performance assessments composed of hands-on performance tasks reported in writing. Boys perform better on multiple-choice tests. One of the plausible explanations provided by Ben-Shakhar and Sinai (1991) is that boys tend to guess more on multiple-choice exams and thus have lower omission rates than girls.

Suppose this were the case in our setting and there were unobserved skill gaps between boys and girls that drove performance gaps for each question type. Consequently, girls could be deemed to perform worse on average in these classrooms, not because the educator holds stereotypes in favor of boys but because the format of the classroom examinations assesses the skills on which they have a relative advantage. Formally, according to Equation (3), this would shift the gender discrepancies in the teacher-assigned scores, reflected in $\beta_{2,j}$, but leave the blindly graded score gaps, $\alpha_{2,j}$, unaffected. I address this concern by examining the distribution of question types per competency evaluated across examinations. As shown in Online Appendix F, I find no substantial gender difference in the distribution of questions nor the distribution of questions per competency assessed in the standardized and classroom examinations.[27] Nevertheless, there might be exam-type-related skill gaps that remain unobserved and can potentially shift $\theta_j$ to the disadvantage of boys or girls.[28] The negative mean of the estimate $\hat{\theta}_j$ could be interpreted as evidence that teachers have an inherent stereotypical view against male students. An alternative explanation is that the mean gender gaps on the teacher-assigned and centrally-assigned tests may be different because they measure slightly different sets of skills. As has been found for GPA versus standardized tests, teachers tend to reward conscientiousness more on in-class teacher assessments than on standardized tests. This would not affect my estimations because I focus on variations in these gaps across educators rather than their absolute size. If there are indeed gender-based differences in test performance, but those differences are uniform across instructors, as required by my identification assumptions, then my gender differences metric can still be used to assess faculty members' systematic favoring of boys or girls.

---

[26]The comparison of examination contents and formats for teacher-graded and blindly graded examinations corresponding to the 2019 school year is contained in Online Appendix F. I chose 2019 as a benchmark year for the comparison because it corresponds to the most recent evaluation year in my analysis period. Appendix Table F.2 contains the comparison of contents in mathematics evaluations, and Appendix Table F.3 displays the analysis of language arts examinations. Moreover, Appendix Table F.4 displays the comparison of contents assessed across different question types.

### 4.2.2 Gender-math stereotypes predict gender differences in assessment

In this section, I examine the extent to which observed teacher characteristics and (mostly) unobserved implicit gender stereotypes are predictors of estimated gender differences in assessment. Consider the following regression relating teacher $j$'s estimated systematic grading differences to their observed characteristics:

$$\hat{\theta}_j = \phi_1 + X'_j \phi_2 + u_j \tag{4}$$

Here, $\hat{\theta}_j$ is the stereotyped grading estimate for teacher $j$ divided by the corresponding subject's bias-corrected standard deviation reported in Table 5. The covariate vector includes demographic characteristics (for example, gender, age), job-appointment characteristics (for example, experience in private and public schools, qualifications), and performance outcomes from national teachers' evaluations as proxies for teaching skills. This specification includes school fixed effects and missing-value dummies where the covariate information is unavailable; standard errors are clustered at the school level. In this exercise, I weight each teacher-level gender differences in assessment coefficient by the inverse of the associated standard error, $s_j^2$, to increase precision.

Similarly, I estimate the following regression equation that relates teacher $j$'s estimated grading gender differences and their implicit gender attitudes as measured by the IAT:

$$\hat{\theta}_j = \gamma_{1,s(j)} + IAT'_j \gamma_2 + X'_j \gamma_3 + \kappa_j \tag{5}$$

Here, the teacher-level gender differences in assessment is divided by the bias-corrected standard deviation. $IAT_j$ denotes the IAT score of teacher $j$, which has been standardized to have mean of zero and standard deviation of one. The index $s(j)$ indicates the school location $s$ where teacher $j$ is currently teaching according to the survey data; thus, $\gamma_{1,s(j)}$ are school-location fixed effects, and $X_j$ is a vector of teacher characteristics that include both demographic and job-related characteristics drawn from administrative records and collected by survey. The covariates include teachers' children status (that is, whether they have daughters, sons, or both), ethnicity indicators, and the decade of birth; the order of the associations in the IAT (that is, whether order compatible or incompatible appears first); and the number of previous IATs the teacher has taken. Observations are at the teacher level, and standard errors are clustered at the school location level.

---

[27]Appendix Table F.4 reports the share of questions from each type of evaluation used to assess each learning competency for eighth grade. Only 5 out of 16 competencies were assessed on teacher-graded tests through a share of questions substantially different from the share used to assess the same competency in the standardized test (that is, share differences between five and nine percentage points).

[28]The second identifying assumption requires the skill differentials between boys and girls to not vary by teacher. This assumption ensures that, even if there are skill gaps across gender, the teacher-specific measure of systematic gender gaps in assessments has a meaningful interpretation for comparisons across exposure to different teachers stereotyped evaluations.

**Gender differences in assessment in relation to teacher characteristics.** Teacher demographics and teaching experience are stronger predictors of assessment-based gender differences than job experience or performance metrics in their jobs. Table 6 summarizes the estimated relations defined in Equation (4) for different configurations of teacher characteristics for the sample of mathematics instructors according to administrative records. The first two columns report coefficients for the entire sample of math teachers. The last column does so for the subsample of teachers who participated in the government's National Teacher Evaluation during 2015–19 in pursuit of tenure-track appointments. Columns (1) to (3) demonstrate that teachers with large systematic gender differences in grading in disfavor of female students tend to be women. This finding contrasts with that of Carlana (2019) and Alesina et al. (2018): female teachers are likely to exhibit less stereotypes and prejudices against girls, as measured by the IAT. I consider these results to document the extent to which gender stereotypes have been internalized by female teachers in my setting, in accordance with the discussion in Section 2.

In addition, my results show that being older than the median age (45 years old) in this sample of teachers is associated with a weaker grading gender differences.[29] There are two possible arguments to explain this finding. One is in line with Altonji and Pierret's (2001) work that posits that prior prejudiced attitudes based on group membership (for example, race, gender) can be updated with continuous interactions with members of such groups. However, there is also a vast literature suggesting strong cohort effects on reported beliefs regarding discrimination.[30] In the Peruvian context, teachers' self-reported gender attitudes collected through my nationwide representative survey suggest there are no substantial cohort effects. Appendix Figure C.5 shows that, for the sample of surveyed mathematics teachers matched with a gender differences in grading measure, there is no marked relationship between teachers' age and gender attitudes across six self-reported measures. Nevertheless, analyzing the evolution of these gender attitudes is impossible with the available information.

Columns (2) and (3) provide evidence that more experienced teachers with private school backgrounds tend to exhibit weaker gender differences in assessments, while previous employment in public schools has less bearing in predicting assessment gender gaps. The pattern is consistent with Alan et al.'s (2018) finding that teachers with modern teaching styles are less likely to hold traditional beliefs about gender (including a warm approach and a growth mindset). In my empirical setting, teachers who opt to teach in, and have more experience in, private schools often adopt state-of-the-art teaching styles. Qualifications are similarly essential, as university-educated instructors tend to exhibit smaller gender-based disparities in grading across all specifications. Column (3) indicates that among the evaluated teachers, those who performed better on a knowledge-based test and were approved for a nationwide round of evaluation usually exhibit smaller discrepancies in assessments of boys and girls.[31]

---

[29]The median age in the sample of teachers reported in Columns (1) and (2) was 45 years in 2019, while for the sample of evaluated teachers in Column (3), the median age was 40 in 2019. However, I maintained the same definition of median age at 45 years across all columns in the table.

[30]In the US, in the last two decades racial attitudes have been declining overall, which has been attributed to time and cohort effects (Charles and Guryan, 2008).

Teachers' characteristics and qualifications are critical to comprehend to what extent the estimated systematic gaps in assigned scores between girls and boys actually reflect discriminatory behavior based on gender. However, it should not be understood as the driving force of gender stereotypes. Instead, the estimated relationships provide further support regarding the direction of the discrimination: higher values of $\hat{\theta}_j$ correctly identify teachers stereotyped assessments against female students.

**Stereotyped assessments: relationship between gender stereotypes and teachers' grading gaps.** This study presents crucial new evidence on the extent to which implicit stereotypes, measured by IAT scores, undergird the observed gender differences-in-assessment estimates, $\hat{\theta}_j$. I leveraged survey data and matched a subsample of 2,052 mathematics and language arts teachers with IAT measures and the estimated stereotyped assessment measure. The distribution of the IAT scores for teachers in this sample is displayed in Figure 3. The left figure shows that male mathematics teachers have higher IAT scores across the distribution, while the right figure shows that female language arts teachers have higher IAT scores across the distribution. This indicates that male math teachers associate men with science more strongly than female math teachers, and vice versa for language arts instructors. In fact, the average IAT score in math is 0.301, and among language arts teachers, the average is close, at 0.28. TThis finding suggests that the average Peruvian high school teacher has an IAT score that nearly places them in the highest category of gender stereotypes. Furthermore, gender stereotypes are ubiquitous in the two primary subjects of instruction.

I use gender stereotype survey results to fit the specification in Equation (5). Each observation represents a teacher, the IAT score is the variable of interest, and robust standard errors are clustered at the school location level. Table 7 delivers evidence on the relationship between implicit stereotypes and discrepancies in grading gaps between female and male students. First, I analyze the gender-science IAT's relationship with gender differences in assessments.[32] Panel A reports coefficients in a specification containing only school-location fixed effects. Panel B controls for a set of covariates including teachers' gender, childbearing status, ethnicity, birth decade, the order of associations in the version of the IAT that they took (i.e., whether humanities appeared on the left or right of the screen at the beginning of the test), and the number of times they have taken IATs previously. To the best of my knowledge, this is the first study that establishes that implicit gender stereotypes are a robust predictor of observable systematic

---

[31]The Ministry of Education indicates that between 2014 to 2019, about 30 nationwide evaluations were implemented, allowing more than 400,000 teachers to be evaluated Minedu (2015).

[32]The IAT scores have been standardized for this exercise.

differences in the assessment of girls and boys.

Column (1) indicates that math teachers who associate males more strongly with science and females more strongly with humanities on the IAT are more likely than other teachers to exhibit larger discrepancies between teacher- and blindly-graded evaluations, to the disadvantage of girls. In other words, teachers who think boys are better than girls in math- and science-related fields give boys better grades than they deserve in mathematics, whereas they give girls lower grades than they deserve. This relationship prevails whether or not I include covariates. In contrast, column (2) indicates that language arts teachers who strongly associate males with science and females with the humanities are less likely to create large score gaps in language arts that disadvantage female students. These coefficients suggest that teachers penalize female students in mathematics but not in language arts since they award higher grades in the latter subject or knowledge domain, where they perceive girls to be more competent or to have greater potential. The direction of the coefficients reported in Table 7 is consistent with subject-specific teachers' gender stereotypes, providing strong evidence that the measurement of systematic differences in grading is a reliable indicator of gender-math stereotypes.

As a final step, I examine whether or not a single teacher's stereotypical grading practices vary significantly between different subject areas, in this case, mathematics, science, and language arts. We might expect that a more stereotyped teacher who holds prejudices against women in math-science will exhibit stereotyped grading against girls in mathematics or science and, to some degree, in other subjects such as language arts as well. I compute OLS correlations between teacher $j's$ stereotypical grading of female students across all possible subject combinations in which she instructs. High school teachers in Peru typically specialize in teaching just one subject, so the sample size is low for this exercise. Appendix Figure C.6 shows coefficients measuring the relation between teacher-level assessment gaps by subject, where the unit of analysis is the teacher. Further evidence that the teacher-level measure of stereotyped grading is internally consistent is found in the positive and statistically significant relationship between stereotyped grading across all combinations of subjects of instruction.

## 4.3 Estimating the distribution of teachers' stereotyped assessments

In this section, I describe the empirical Bayes method I use to correct sampling error in the set of estimates of $\hat{\theta}_j$. The resulting posterior means estimates serve to construct the primary variable of interest in the next step of my empirical analysis.

I leverage the set of unbiased but noisy estimates, $\hat{\theta}_j$, of the true underlying teacher-level stereotyped assessments parameters, $\theta_j$, using empirical Bayes (EB) methods. These methods are particularly advantageous, as they provide estimators that use the observed estimates $\hat{\theta}_j$ for calculating the aspects of the population distribution of the $\theta_j$—beyond mean and variance—that might be of interest, even though the latter values are not themselves observed (Efron and Morris, 1972, 1975, Stein, 1964). Estimators in this framework have desired properties such as minimizing Bayes risk and minimizing quadratic error loss (Efron and Morris, 1972). Therefore, in this section I report precise estimates computed using two EB methods: parametric EB (also known as linear shrinkage estimation; see, for example, Chetty et al. (2014b), Gilraine et al. (2020), Kane and Staiger (2008)) and EB deconvolution (as proposed by Efron (2010, 2012, 2016)). I use these posterior mean estimators to construct the dependent variables in my analysis of the long-term effects of stereotyped grading on long-term outcomes.

### 4.3.1 Parametric empirical Bayes estimator

A natural way to model this problem is hierarchically by interpreting the teacher-level stereotyped grading estimates to be a random sample from a common *prior* population distribution of $\theta_j$. I assume that the population parameters $\theta_j$ are drawn from a normal prior distribution with hyperparameters $(\mu, \phi)$ so that $\theta_j$'s are conditionally independent given $(\mu, \phi)$:

$$\theta_j | \mu, \phi \overset{ind}{\sim} \mathcal{N}(\mu, \phi^2) \qquad , \text{for } j = 1, \ldots, J \qquad (6)$$

Next, consider $J$ experiments (that is, classroom assignments) in which teacher assignment $j$ estimates the parameter $\theta_j$ from a class size $N(j)$ that is independently distributed. This is written

$$\hat{\theta}_j | \theta_j \overset{ind}{\sim} \mathcal{N}(\theta_j, s_j^2) \qquad (7)$$

for $j = 1, \ldots, J$, where $\sigma_\theta^2$ denotes the variance of $\hat{\theta}_j$. I am interested in the joint conditional posterior distribution $p(\theta | \mu, \phi, \hat{\theta})$ of the underlying parameters $\theta_j$, which in this case has the form

$$\theta_j | \mu, \phi, \hat{\theta} \sim \mathcal{N}(\hat{\theta}_j^*, \Sigma_j) \qquad (8)$$

and the posterior mean estimator is

$$\hat{\theta}_j^* = (\tfrac{1}{\sigma_j^2} \hat{\theta}_j + \tfrac{1}{\phi^2} \mu) / (\tfrac{1}{\sigma_j^2} + \tfrac{1}{\phi^2}), \qquad (9)$$

with $\sigma_j^2 = \sigma_\theta^2 / N(j)$ as the sampling variance.[33] A feasible version of this posterior mean estimator only requires the standard error of the associated $\hat{\theta}_j$. Therefore, under the previous assumptions, the parametric EB is a precision-weighted average of the prior population mean and the estimator, $\hat{\theta}_j$, that shrinks the estimated stereotypical grading measure toward the respective sample mean. Appendix Figure C.7 plots the posterior means' distribution under linear shrinkage. These values show lower variability as they are shrunk toward the sample mean.[34]

### 4.3.2 Application of empirical Bayes deconvolution methods

A key assumption of the parametric EB setup is that the prior distribution of the true parameter $\theta_j$ is Gaussian, which implies the posterior distribution specified in Equation (8) also is. The linear-shrinkage methods also carry features regarding the shrinkage procedure proposed in Equation (9). As Gilraine et al. (2020) indicates, the factor of such shrinkage only depends on the sign and not its magnitude. This property is particularly relevant in this setting, as the stereotyped assessment against girls and boys located at opposite sides of the distribution (that is, right tail and left tail, respectively) will be shrunk toward zero by the same factor as long as the class size and estimated stereotyped grading are the same. To assess the importance of the normality assumption, I also estimate a more flexible model for the prior distribution of $\theta_j$. In particular, I follow the modeling approach of the prior density proposed by Efron (2016), which uses an exponential family of densities to model the prior, because of its efficiency compared to other deconvolution approaches.[35] In this setting, I assume an unknown prior density, $g(\mu)$, yielding population parameters distributed as follows:

$$\theta_j \stackrel{ind}{\sim} G(\mu), \text{for } j = 1, \ldots, J \tag{10}$$

---

[33] It is easy to verify that $\Sigma_j = (\tfrac{1}{\sigma_j^2} + \tfrac{1}{\phi^2})^{-1}$.

[34] An issue not directly addressed here is the independence assumption of the estimated $\hat{\theta}_j$. Gilraine et al. (2020) proposes a test to formally test this assumption.

Here, each $\theta_j$ draws independently an observed $\hat{\theta}_j$ following a known exponential probability-density family $f_j$ such that

$$\hat{\theta}_j \stackrel{ind}{\sim} f_j(\hat{\theta}_j|\theta_j). \tag{11}$$

Figure 4 displays the deconvolved density $\hat{g}(.)$ of the teacher-level stereotyped assessment parameters $\theta_j$. I also report the theoretical Gaussian density of the estimated teacher-level stereotyped assessment and its observed distribution. Following Kline et al. (2021), I choose the penalization parameter such that the deconvolved density is calibrated in mean and variance using the bias-corrected variance estimates reported in Table 5. Visual inspection of deconvolved density indicates that Gaussian density is a good approximation for the prior density $g(.)$. These results suggest that a linear shrinkage estimator approximates the posterior means as if they were computed using the true underlying population.

## 5 Empirical strategy for estimating the long-term effects of teachers' stereotypical assessments

### 5.1 Research design

Because students have different teachers throughout their academic careers, their exposure to stereotyped assessments also varies. Let $\hat{\theta}^*_{j,-i}$ denote the posterior mean of the leave-one-year-out estimator of stereotyped evaluations of teacher $j$ of student $i$. This estimator was normalized using its mean and standard deviation calculated per school year.[36]

The main estimating equation to retrieve the causal relationship between teachers' stereotypical evaluations and student $i$'s long-term outcomes variable, $Y_i$, is as follows,

$$Y_i = \delta_0 + \delta_1 \hat{\theta}^*_{j,-i} \cdot Female_i + \delta_2 Female_i + \delta_3 \hat{\theta}^*_{j,-i} + \delta_4 \mathbf{X}_i + u_i \tag{12}$$

The parameter $\delta_1$ captures the differential effects of a one standard deviation increase in teacher stereotypical assessments against girls on girls versus boys. The parameter $\delta_3$ represents the effects of male students being exposed to stereotyped assessments in the classroom. $Female_i$

---

[35]Various methods assess theoretically and empirically the deconvolution problem (for example, Fan (1991)). They can be grouped in g-modeling approaches when the prior distribution aims to be estimated Efron (2016) and in f-modeling approaches when the distribution of the observed parameters is retrieved (see, for instance, Carroll and Hall (1988), Efron (2014)).

[36]The standard deviation of teacher-level stereotypical assessment measure used for standardizing $\hat{\theta}^*_{j,-i}$ are calculated as the bias-corrected measures reported in Table 5 for the corresponding school year.

specifies whether the student $i$ is female.

The covariate vector $\mathbf{X}_i$ consists of four groups of controls. First, I include student-level controls, including students' language, age in months, mother's education, and indicator variables for repetition and place of birth, to account for students' social norms.[37] This set of covariates also includes teachers' characteristics such as gender, age, type of appointment indicators (such as homeroom teacher, subject teacher, and administrative teacher), and, as a proxy for teacher experience, contract length (for example, tenured or fixed-term contract). Depending on the gender of the students, teachers' behavior may vary. To account for the gender-based differences in the effects of educational inputs on human capital decisions, the set of teacher characteristics interacts with student gender.

To account for the compositional effects of assigned classes, I also control for classroom-level characteristics, including class size, and student-level characteristics averaged at the classroom level. The third group of controls consists of school-grade averages of students' characteristics in order to account for confounding variables associated with exposure to stereotyped teachers across grades within the same school. I also include polynomials of classroom and school-grade means for these scores. I also include cohort, grade, school year, and school fixed effects when estimating this model.

**Leave-one-year-out estimate of stereotyped assessments.** My preferred specification uses a leave-one-year-out version of the teacher-level stereotyped assessments metric, denoted $\hat{\theta}^*_{j,-i}$, which is computed using the posterior-mean estimates introduced in Section 4.3. The leave-out method avoids any potential correlation between the sampling variance in $\hat{\theta}_j$ and long-term outcomes that could be introduced by using the same students to study teacher stereotyped assessments and long-run outcomes. For example, consider the computation of the leave-one-year-out teacher's stereotypical grading, which is the treatment variable for students expected to graduate in 2019. I exclude the data from all of the 2019 graduation cohort examinations (that is, data on eighth-grade students who took the examinations in 2015) and use the remaining students' observations to calculate all teachers' stereotyped grading. I repeat this exercise for students in all other (projected) graduating cohorts.[38] This calculation results in a leave-one-year-out teacher-level assessment stereotype estimate for each student cohort based on their projected graduation year.

## 5.2 Identification of long-term effects

Students may not be randomly assigned to teachers' classrooms, which is a significant concern when estimating the effects of exposure to stereotypical grading practices. Consequently, the estimated effects of teachers' stereotyped assessments may not exclusively capture the effects of their gender stereotypes on students' decisions (as reflected in student outcomes), but also the influence of other gender-specific unobservables that systematically drive these educational investments, such as ability differentials reinforced by gendered social norms.

---

[37]IIn accordance with Carlana (2019), I consider student social norms to be related to the place of birth; thus, I include birthplace indicators as covariates. Sociological studies also document this relationship. For example, see Crawley et al. (2013), Uunk and Lersch (2019).

[38]Only the cohorts with projected graduations in 2018 and 2019 took the standardized examinations in 2015 and 2016, respectively. I construct a leave-one-year-out measure of teachers' stereotypical grading, excluding 2015 and 2016 scores, and assign them exclusively to these cohorts. The remaining cohorts are assigned the regular teacher-stereotyped assessment estimates that combine information from 2015, 2016, 2018, and 2019 scores.

To formally examine the potential confounding variables that prevent me from isolating teachers' stereotyped assessments, I build on the potential-outcomes framework.[39] Consider that the variable for potential long-term outcomes, $\tilde{Y}_{ij}$, for student $i$ assigned to teacher $j$ depends on the scores given by the teacher and central planner:

$$\tilde{Y}_{ij} = f_i(X_i, \tilde{S}_{ij}^B, \tilde{S}_{ij}^T, \rho_i) \tag{13}$$

For instance, $\tilde{Y}_{ij}$ might refer to high school completion. Here, $\rho_i$ is a composite of students' unobserved characteristics (for instance, their educational aspirations or prior beliefs regarding their academic potential) while $X_i$ is a set of student-demographic characteristics affecting academic-progress decisions. Moreover, student $i$'s centrally assigned score if they were assigned to teacher $j$ is defined as follows:

$$\tilde{S}_{ij}^B = s_i(M_i, W_{ij}, \psi_i, \eta_{ij}) \tag{14}$$

$M_i$ is an indicator variable denoting the membership of student $i$ in one of the two classroom groups (for example, boys). This specification allows for group membership to directly affect centrally assigned test scores through factors other than stereotypes. $W_{ij}$ is a covariate vector of characteristics from student $i$ and their assigned teacher $j$. $\psi_i$ represents the innate ability or academic potential of the student. The score assigned by the teacher $j$ if assigned to student $i$ is defined as follows:

$$\tilde{S}_{ij}^T = c_i(M_i, W_{ij}, \psi_i, \vartheta_{j,g(i)}, \eta'_{ij}) \tag{15}$$

The parameter $\vartheta_{j,g(i)} \in \mathbb{R}$ captures teacher $j$'s forecast of student $i$'s ability based on the student's group membership $g$. We can think of this unobserved parameter as corresponding to teachers' stereotypical assessments against a given group if the prior beliefs of any student in that group are systematically lower.[40] Unlike in the previous equation, in which a blind examiner assigns the score, here I allow the examiner to idiosyncratically forecast student $i$'s ability based on the student's group membership $g(i)$. Both $\eta_{ij}$ and $\eta'_{ij}$ are unobserved determinants of scores and are orthogonal to all other factors in their respective equations. I assume they have means of zero and are identically distributed and pairwise independent for any pair $(i, j)$.

I now consider how the human capital investment decision depends on on potential scores assigned by the teacher and central planner:

$$\tilde{Y}_{ij} = f_i(X_i, M_i, W_{ij}, \psi_i, \vartheta_{j,g(i)}, \tilde{\eta}_{ij}) \tag{16}$$

Here, $\tilde{\eta}_{ij}$ combines any unobserved gender-driven difference in student or teacher characteristics, other than stereotypical teacher assessments, that influences the outcome with any unobserved gender-related characteristic that influences the outcome. Among the first category of unobservables, for instance, there could be parents' requests to assign their children to a teacher who shares their progressive views on educational equality. In the latter group of unobservables, high-performing students are more likely to be assigned to nonstereotypical teachers who do not interfere with their self-assessment of academic potential.

---

[39] See Online Appendix Section B.2 within a dynamic human capital investment model. In that setting, I allow students to have asymmetric information on their abilities or academic potential and teachers to have gender-stereotyped assessments and provide performance feedback that guides pupils' decisions. In addition, centrally assigned scores are given as signals for students' decisions.

[40] A formal definition is introduced in Definition 1 in Online Appendix Section B.2.

In light of these concerns, the following equation expresses the selection-on-observables assumption from which I derive the causal estimates of being exposed to a teacher who makes more stereotypical assessments of any student based on their group membership:

$$\mathbb{E}[\tilde{Y}_{ij}|\hat{\theta}^*_{j,-i}, M_i, \mathbf{X}_i, j(i) = j] = \lambda_0 + \lambda_1 M_i + \lambda_2 \hat{\theta}^*_{j,-i} + \lambda_3 \mathbf{X}_i; \qquad j = 1 \ldots J \qquad (17)$$

Here, $\mathbf{X}_i \equiv [X_i, W_{ij}]$ is a covariate vector of teacher characteristics, student characteristics, and lagged scores, and $\hat{\theta}^*_{j,-i}$ is the empirical counterpart of the parameter introduced in Equation (3). In other words, a student's assignment to teacher $j$ is assumed to be orthogonal to unobserved determinants in levels and unobserved differences across genders that guide the decisions regarding high school completion, college outcomes, and labor market outcomes, after conditioning on observed characteristics of students, teachers, classroom-level and school-grade-level characteristics, and lagged student scores. This selection-on-observables assumption is often proposed in the value-added literature to alleviate selection concerns driven by students' sorting mainly on ability (see Chetty et al. (2014a,b), Kane and Staiger (2008), Rothstein (2010, 2017)). In this case, the conditioning variables include observable characteristics of the student, $\mathbf{X}_i$, that directly determine their investment in qualifications as well as other characteristics $M_i, W_{ij}$—such as educational inputs, students' group membership, and school and teacher characteristics—that influence students' decisions regarding whether to continue academically.

A further consideration for identifying the underlying causal parameter is that the estimated teacher-level stereotypical grading is computed using the scores of cohorts of eighth-grade students from a particular school year. As long as it is displayed in the same subject, I assume that a teacher's stereotypical grading of previous years and specific cohort-grades is predictive of the stereotypical grading of future cohorts in other grades. This is consistent with the value-added literature, which measures teachers' or schools' value added based on the scores of students from prior periods and other student cohorts (Abdulkadiroğlu et al., 2020, Rothstein, 2017).

## 6 Results on academic progress and college outcomes results

### 6.1 Effects on high school completion

My first main finding is that female students, who are as good as randomly assigned to math teachers with higher stereotyped assessments against girls in math-science, are less likely to graduate high school on time or ever. Table 8 shows the effects of exposure to a one-standard-deviation mathematics teacher with more stereotypical assessments against girls in math-science, during one grade, on the likelihood of graduating high school. The outcome variable in Columns (1) to (4) is an indicator variable equal to 1 if the student graduated in the school year corresponding to their projected high school graduation date. Columns (5) to (8) have as outcome variables an indicator of whether the student graduated a calendar year after their projected graduation date. The window of time over which graduating from high school *ever* is defined varies from cohort to cohort, ranging from one calendar year after projected graduation for the youngest cohort (class of 2019) to four for the oldest cohort (class of 2015).

Exposure to a more stereotyped teacher increases the gender gap in high school graduation. This can be seen in Table 8, which reports estimates of Equation (12) separately by grade. In Column (1), the interaction coefficient indicates that a one-standard-deviation increase in stereotypical grading practices during eighth grade lowers the likelihood of graduating from high school by 3.6 percentage points for girls relative to boys. The following columns suggest that exposure to a more stereotyped teacher in ninth grade has a differential effect on girls of 1.7 percentage points, while if exposed during tenth grade, girls are 1.5 percentage points less likely to graduate relative to boys. Overall, the point estimates follow a slightly decreasing trend over time as the students advance to senior grades.

These grade-specific estimates are aggregated into a single summary effect in Column (4). This specification stacks the data for all high school grades, with standard errors clustered by student. On average, exposure to a more stereotypical math teacher for one grade during high school, to the detriment of girls, widens the high school graduation gap by 1.5 percentage points (1.8% of the mean). Columns (5) through (8) repeat this analysis with an outcome variable equal to one if the student ever graduated from high school, including late graduation. The summary of the points estimates is reported in Column (8). In this case, a one-standard-deviation increase in stereotyping is estimated to increase the gender gap in graduation by 1.3 percentage points (1.6% of the mean). The exposure of boys to a teacher with one standard deviation stronger evaluation stereotypes increases their likelihood of graduating by 0.6 percentage points (0.7% of the mean).

### 6.2 Effects on college attendance

This section discusses the causal effects of teachers' gender stereotyped assessments on college outcomes, focusing on enrollment in four-year colleges rather than two-year or vocational schools. Students' average age at graduation is 16 years, and most students decide to apply to college in their junior year of high school (that is, tenth grade).[41]

The persistence of teachers' stereotyped assessment effects negatively impacts the college application gender gap. Table 9 presents the estimates of exposure to teachers' stereotypical grading once the students have graduated from high school and begun college. Columns (1) and (2) show a statistically significant impact on the gender gap in college applications. The estimates indicate that a one-standard-deviation increase in stereotypical grading reduces girls' likelihood to apply to college on time by 0.6 percentage points relative to boys. Similarly, a one-standard-deviation increase in stereotyped grading lowers the likelihood that girls will ever apply to college by 0.7 percentage points relative to boys (2.2% of mean). These results differ from the findings of a recent study by Lavy and Megalokonomou (2019), in which exposure to teachers exhibiting more stereotypical grading against girls in the last two high school grades decreases the likelihood of postsecondary enrollment for girls and increases it for boys. My results indicate that the exposure of boys to stereotypical grading has no statistically significant effect on their likelihood of applying to college. According to the estimates in Columns (3) and (4), teacher stereotypical assessments have no significant effect on the gender gap in college admissions. In the last two columns, enrollment in college yields a similar result. In combination with the results for high school graduation, these findings imply that the primary impact of teacher stereotypes on educational attainment occurs at the margin of high school graduation.

---

[41]Online Appendix Section A.1.3 provides a brief description of the university system in the country.

The evolution of the effects of grade-specific exposure to teachers' stereotyped grading on college applications draws a detailed picture of how relevant it in junior year (eleventh grade) and freshman year (eighth grade). Figure 5 plots the main effects of exposure to teacher-level stereotypical assessments against girls on the likelihood of applying to college, as well as the interaction effects for girls by high school grade. I find negative and statistically significant effects on the gender gap in college application during the eighth and eleventh grades. The large effects of teacher stereotyped grading found during eighth grade may be a result of the performance signal sent to parents and teachers based on students' performance on national standardized tests. Parents receive a comprehensive report comparing their child's performance in mathematics, language arts, and science to the national average and the performance of students by gender. Parents are informed whether their children's achievement levels are "Satisfactory," "In the process," "Beginning," or "Before beginning." In addition, teachers receive a comprehensive report that ranks students within the class and compares their performance to that of other eighth-graders in the region and the national average.

### 6.3 Robustness checks on high school completion and college outcomes

The empirical exercise aims to identify two parameters introduced in Equation (12). First, the interaction parameter $\delta_1$ captures the differential effects on outcomes among girls exposed to teachers whose stereotypical grading practices exceed the mean by one standard deviation. Second, the parameter $\delta_3$ is interpreted as the effect on male students' outcomes of being exposed to a teacher with one standard deviation higher stereotypical grading than the average teacher. The first potential threat to identification arises if students are sorted to more or less stereotypical assessment teachers based on observed or unobserved characteristics. I address this concern by showing in Appendix Table C.2 that students' and teachers' baseline characteristics that potentially drive such assignments are not systematically related to the teacher-level stereotypical grading measure.

The second threat stems from the possibility that teachers' stereotypical evaluations of eighth-grade students may differ from their stereotypical evaluations of students in higher grades. This could occur if, for example, principals assign more conservative and stereotypical assessment teachers to teach the upper grades of high school. To determine whether such a scenario is of concern, I calculate the long-term effects on a restricted sample consisting of only the graduating classes of 2018 and 2019, which can be observed in both lower and upper grades. The results reported in Appendix Table C.3 to Appendix Table C.5 show that the estimates of the effects of math teachers' stereotyped assessments on graduation outcomes remain statistically significant in the subsample of students described. Furthermore, find statistically significant effects on college outcomes for the restricted sample, as displayed in Appendix Table C.5.

Finally, there is also the possibility of unobserved patterns in teacher assignments to classrooms across grades, such that the absence of stereotypical assessment measures for some grades is not random. To examine this potential threat, I define a subsample consisting of all student cohorts whose exposure to teachers' stereotypical grading is observed throughout all grades 7 to 11–that is, for every grade they appear in the sample. The results reported in Appendix Tables C.6 to C.8 do not indicate that this is a cause for concern.

### 6.4 Teachers' stereotyped assessments versus value added

As discussed in Section 2, teachers' value added is a channel that could potentially mediate the effects I find on academic progression and college outcomes. Therefore, Consequently, I investigate whether teachers exhibit a differential value added toward boys and girls in their classrooms. These gender-specific value-added estimates allow me to investigate whether higher-quality teachers exhibit more or less stereotypical assessments.[42]

To formally analyze this mechanism, consider the estimating Equations (1) and (2), which form a system. I am interested in retrieving the following parameters. First, I define $\alpha_{1,j}$ as teacher $j$'s value added toward girls and $\alpha_{1,j} + \alpha_{2,j}$ as teacher $j$'s value added toward boys. I compute these parameters using a fixed-effects–SURE specification according to the previous equations, allowing for error terms to be correlated across equations.

With estimates of teacher value-added and stereotyped grading in hand, I calculate the joint covariance matrix of these parameters. This intermediate step is needed for conducting a bias correction using the sampling covariances calculated from SURE estimation. Bias-corrected estimates of the correlations in value-added and bias parameters across teachers appear in Table 10 (see details on the estimation procedure in Online Appendix E). Consistent with Lavy and Megalokonomou (2019), I find that teachers with more stereotypical grading practices in mathematics have lower value-added scores for both girls and boys. Consistently with Table 5, I find a lower dispersion of stereotyped grading among mathematics teachers than among language arts teachers under this specification. The estimated correlations in Columns (2) and (3) reported in Panel A indicate a stronger negative association with value added toward boys. The same patterns are present in Panel B among language arts teachers. This could reflect ripple effects on the classroom climate when teachers are more stereotyped against girls (Bertrand and Pan, 2013, Figlio et al., 2019). In addition, between the value added for boys and girls in math and language arts, which contradicts the gender-based match effects found in recent research (Dee, 2005, Gershenson et al., 2018). This analysis suggests that low-value-added teachers tend to have more stereotypes. Consequently, teachers' quality is an essential mechanism by which stereotypical assessments reduce high school completion and college applications.[43]

---

[42]To complement this section, Appendix Table C.14 reports the effects of teachers' stereotyped assessments on gender score gaps in mathematics and language. The effects of teachers' gender stereotypes are more pronounced in the grades of the subjects they teach. For example, mathematics teachers have a greater impact on math score gaps than they do on language arts score gaps.

[43]I also investigated the tracking of students to stereotyped teachers across grade levels in Appendix Table C.9. I found a positive correlation between stereotyped assessments in one grade and the following grade. This is however explained by repeated assignments to the same teacher, as the correlation becomes negative if the sample is restricted to students who did not have the same teacher in consecutive years.

## 7 Effects on labor market outcomes

### 7.1 Employment in the formal sector

This section reports the main findings on the set of labor market outcomes for the students' cohorts graduating between 2015 and 2019, whose 18-to-23-year-old trajectories I follow (the former is the minimum legal age). The regression sample allows me to analyze the effects of stereotypical teacher assessments on a sample of high school graduates who either have full-time jobs in the formal sector or work part-time while attending higher education institutions. Given the age of the workers, the majority of the students' positions are non-professional entry-level positions. My results indicate that the increased exposure of female students to stereotypical grading by teachers has a significant and lasting impact on the formal sector employment of girls aged 18 to 23.[44]

I examine the effects of exposure to a more stereotypical grading teacher using the research design proposed in Section 5. n this section's specification, I account for social norms-related local labor market dynamics by interacting the student's place of birth with their gender. Table 11 display the estimated effects on the likelihood of formal sector employment from the empirical model in Equation (12). Each column indicates the modal ages of students throughout their first three years in the labor market. Column (1) reports the estimated effects when students in the sample have reached the minimum legal working age of 18 years. The effects of teachers' stereotypical assessments on the gender gap in formal employment are 0.1 percentage points (18% of the mean). The bottom rows of the table report the mean outcome values for female and male students, indicating that approximately 93 percent of 18–19-year-old girls in the sample are likely exmployed in the informal sector. Moreover, the detrimental effects on the gender gap in formal employment are 2.4 percentage points at ages 22–23 (21% of the mean), as shown in Column (3). This is a large effect equal to 27 percent of the baseline gender gap and 18 % of the average formal sector employment rate.

Figure 6 summarizes the evolution of teachers' stereotypical assessment effects by displaying coefficients analogous to those in Table 11. The lines display the estimated main effects of teachers' stereotyped grading, the interaction effects, and the gender gap in formal sector employment; the confidence intervals are displayed in the shaded regions. The differential effects on girls remain negative and persistent until they reach 22–23 years of age. Moreover, the total stereotyped grading effect on female students is negative and statistically significant between the ages of 18–19 and 20–21, ranging between -0.08 percentage points and -0.64 percentage points, respectively. In contrast, the total effect on male students is positive across all years: is 1.18 at ages 18–19 and 2.51 percentage points at ages 20–21.

Two main features of the labor market in the country support the salient total and differential effects that I find. On the one hand, the size of the informal labor sector in the country is considerable and vastly absorbs young workers. Between 2015 and 2019, the average share of informal jobs nationwide is 64 percent, above most of the countries in the region (International Monetary Fund, 2022). Informal sector jobs in Peru do not comply with labor law stipulating the right to have a contract, a retirement pension fund, health insurance, or unemployment insurance. On the other hand, another distinct characteristic of the labor markets in Peru is that the educational attainment of the labor force is underdeveloped: 68 percent of workers in the labor force have a high school diploma or less.[45] In this context, female high school graduates without job experience find it more challenging to obtain any position in the formal sector: the share of female workers under 29 years old is the lowest among workers with formal jobs (10 percent in 2018). In sum, attaining formal sector employment is highly competitive for high school graduates, given that they are unable to compensate for their lack of employment experience with qualifications that differentiate them from the rest of the workforce. My results suggest that encountering stereotyped educators during high school will likely sidetrack students into detrimental employment trajectories in the informal sector.

---

[44]In Appendix Table C.15, I report the effects of teachers' stereotypical grading on students' employment outcomes between ages 14 and 17 who are considered underage children with legal impediments preventing them from working in any sector, including the formal sector. The minimum legal age to work in the country is 18, so hiring companies will likely require a written written permission from parents or legal authorities for these students to obtain legal employment contracts in the formal sector. Therefore, I concentrate on the labor outcomes of students beginning at the age of 18, which reflect decisions made without parental influence.

## 7.2 Paid working hours

I also find evidence of strong and large effects of stereotyped grading on paid monthly work hours in the formal sector, a result that appears in Columns (1) to (3) in Table 12. Column (1) indicates adverse effects among female graduates ages 18–19. Increased exposure to a one-standard-deviation more stereotypical assessment teacher during one grade year of high school reduces women's monthly paid work hours by approximately one hour. This differential effect accounts for approximately 28% of the gender gap in paid work hours, but it appears to diminish over time. In Columns (2) and (3), the unfavorable effects on women are no longer statistically significant. Similarly to the effects found on the likelihood of holding a formal sector job, the total effects on women's paid working hours (the sum of the first and second rows) are no longer unfavorable to women aged 22–23. However, there is no clear pattern of convergence in the number of paid hours worked by men, who work nearly twice as many hours as women between the ages of 22 and 24.

## 7.3 Monthly earnings

I now present estimates of the monthly earnings losses for recent high school graduates between the ages of 18 and 23. The point estimates in Columns (4) to (6) in Table 12 are expressed in 2010 USD. The first row indicates that women experience persistent effects that aggravate the wage gap between mean and women between the ages of 18–19. My estimates in Column (4) indicate that for women relative to men, the monthly earnings loss caused by having a teacher who is one standard deviation more stereotypical during a single grade is USD 2.6 per month at this age –roughly three years after graduating high school. The baseline monthly earnings gap in the sample by the time the students have 18–19 years is equivalent to USD 8.3 per month; therefore, my findings indicate that exposure to stereotyped grading exacerbates the earnings gender gap by approximately one-third. In addition, the point estimate in Column (4) indicates that the monthly loss in earnings for women exposed to a teacher who is one standard deviation more stereotypical in her grading is equivalent to 1% of the monthly minimum wage at ages 18–19. During the period of analysis, 2025–2019, the monthly minimum wage averages out to about USD 240.

---

[45]According to the National Institute of Statistics and Informatics of Peru, in 2019, 45 percent of the working-age population had a high school diploma as the maximum educational level attained, while 24 percent had elementary schooling as the educational level attained.

Moreover, although the effects at ages 20–21 and 22–23 reported in Columns (5) and (6) are not statistically significant, they have economic importance. A potential cause for teachers' stereotypical grading effects on women to dissipate at this stage is that women have gained significant on-the-job experience to offset the low investments endured during high school. Also, work-study employees might have passed a milestone in higher education degrees that allows detrimental effects to be less significant. In fact, by the time students are 20–21, those enrolled in a university or technical higher education have accumulated enough academic credits to start paid apprenticeships that are regulated by law and overseen by their respective higher education institutions.

Turning to the total effects on male (second row) and female graduates (sum of the first and second rows), Columns (4) to (6) in Table 12 show that the total effects on women are adverse starting at ages 18–19 and fade out only when students are aged 22–23. Therefore, any earnings catch-up by women starts five years after high school, leaving most of them with a long-lived comparative disadvantage in labor markets compared to male graduates. In contrast, exposure to a more stereotyped teacher leads to earning-gains patterns for men aged 18 to 23 —that is, five years after graduating from high school.

Next, I assess whether the effects on earnings arise from low- or high-paying jobs by estimating the effects of teachers' stereotyped assessments on the probability of attaining a high-paying job. In this exercise, I maintain the same research design proposed in Equation (12), using as an outcome an indicator variable that takes the value of 1 when the student $i$'s monthly earnings are above a given earnings percentile. The percentiles are computed from the population of workers with comparable work experience between ages 16 to 25. Figure 7 plots, for each earning percentile, the effects on the probability of having a job with earnings above that percentile when the workers are ages 18–19. Here, I consider the jobs above the fifty percentile as high-earning jobs. Panel (a) shows the women's differential effects and the baseline gender gap of increased exposure to teachers' stereotypical grading in the likelihood of having a high-earning job. Similarly, Panel (b) shows the main and differential effects. The first Panel demonstrates that girls whose monthly earnings correspond to the bottom fifty percentile are less likely to have high-earning jobs when exposed to a one-standard-deviation more stereotyped teacher. Panel B confirms that the total effect on the likelihood of high-earning jobs among girls in the bottom fifty percentile is negative and persistent at ages 18–19 –three years out of high school.

My results indicate marked effects on the monthly earnings gap deriving from low-value-added stereotyped teachers' that disproportionately deviate female students onto low-paying jobs when employed in the formal sector. A partial explanation of why the effects, as a share of the pay gap, are substantial lies in the weak enforcement of recent legislation stipulating equal pay for women and men.[46] As a result of the weak legal protection –which is common in middle and low-income countries– the gender earnings gap in Peru is substantial both in the formal and informal labor markets. For example, in 2015, the raw monthly earnings gap between male and female workers was 29.9 percent (INEI, 2020), while the adjusted earnings gap in hourly wages was 19 percent in 2015 (Muller and Paz, 2018). Unbinding equal pay rights enforcement places little to no boundaries on earning disparities generated in schools.

---

[46]In 2017, laws to avoid pay discrimination between men and women were passed by the Peruvian government. However, since the law's enactment, the judiciary has started prosecuting pay discrimination in 2021.

Moreover, consistent with the literature on the gender pay gap, the distribution of males and females across occupations and industries is a key explanation for persistent pay differences (Cortes and Pan, 2018, Kunze, 2018). Blau et al. (2012) and Blau and Kahn (2017) emphasized that gender differences leading to occupational sorting are pronounced and play a significant role in the observed pay gaps —especially among non-college-educated women whose lack of qualifications prevents them from equalizing payment with male colleagues. Following this line of work, Goldin (2014) suggests that convergence of payment has been more favorable in occupations at the upper tail of the wage distribution. Less is known about what other factors take relevance in explaining earnings gaps at the lower tail of the wage distribution, where the high school graduates are most likely to be.

Appendix Figure C.8 indicates that industry sorting is a likely channel for educators' influence on earnings trajectories among young low-skill workers in Peru. My research suggests that the longer women spend in the workforce after high school, the more industries in the formal economy they are discouraged from entering. The results indicate that stereotypical grading teachers discourage women while encouraging men to seek employment in industries such as "agriculture, forestry, fishing," "transportation and warehousing," "mining, quarrying, oil and gas extraction" and "real estate and rental and leasing" between ages 18-23. Notably, the figure demonstrates that teachers' stereotypical grading has no effect on the participation of females and males in high-skilled industries such as "finance and insurance,", "educational services," "health care and social assistance," and "'other services" where high-skilled professionals tend to be employed in Peru. Overall, exposure to stereotypical teacher assessments exacerbates the gender pay gap by inducing strong sorting effects among male students and narrower sorting mediation among females.

In conclusion, the prevalence of weak labor rights protection and gender-based occupation sorting creates an environment in which the role of educators in determining human capital and vocational interests can divert students toward low-paying jobs in which they are likely to face substantial wage disparities.

### 7.4 Students' internalization of stereotyped assessments

One potential mechanism through which stereotyped assessment teachers might affect students is changes in the student's own gender views. For example, a stereotypical grading teacher might increase negative gender attitudes among girls; a phenomenon known as "internalized bias," as discussed in Section 2. I study this possibility using IAT scores collected in a sample of a younger cohort of students currently in high school and who are projected to graduate between 2022 and 2025. In particular, I use the gender-science IAT scores of these students. The information for this exercise was collected through remote and in-person sessions according to the timeline displayed in Figure 2. Focusing on the in-person sessions, approximately 4,000 students from 266 classrooms from eighth to eleventh grade participated in the data-collection sessions. The student sessions were carefully designed to promote the active participation of students and were carried out by educators hired to lead class-format workshops where students filled out the IAT and a survey.[47] The IAT is a psychological test designed initially for adults, although its validity has also been investigated in children between 7 and 17 years old (Roh et al., 2018). The version that 13–16-year-old students took was adapted from the original version to ensure they comprehended the test's language and contents and stayed engaged while taking the test. Although an educator external to the school was in charge of executing the session activities, these were carried out in the presence of a teacher (neither mathematics nor a language instructor) previously appointed for this task by the school principal.

Using this information, I test whether teachers' implicit gender stereotypes as measured by the IAT, $IAT_j$, are internalized by their students according to the following regression equation.

$$IAT_i = \tilde{\nu}_0 + \tilde{\nu}_1 \cdot IAT_j \, Female_i + \tilde{\nu}_2 Female_i + \tilde{\nu}_3 IAT_j + \tilde{\nu}_4 \mathbf{X}_i + \tilde{e}_i \tag{18}$$

In addition, I estimate the internalization of teachers' gender stereotypes as measured by the assessment-based measure of gender stereotypes and the implicit gender stereotypes measure. A second regression relates student $i$'s gender stereotypes, as measured by the IAT, with the teacher-level stereotypical assessments, $\hat{\theta}^*_{j,-i}$, and maintains the same research design presented in Section 5.1.[48]

$$IAT_i = \nu_0 + \nu_1 \cdot \hat{\theta}^*_{j,-i} \, Female_i + \nu_2 Female_i + \nu_3 \hat{\theta}^*_{j,-i} + \nu_4 \mathbf{X}_i + e_i \tag{19}$$

I find evidence suggesting that exposure to stereotyped teachers can lead female students to internalize the assessment-based gender stereotypes of their educators in mathematics classes.[49] Columns (1) and (2) in Table 13 indicate there is a positive and statistically significant differential effect on girls' own gender stereotypes due to exposure to teachers' implicit gendered assessments. According to these estimates, assigning a mathematics teacher with one standard deviation above the average teacher's IAT increases female pupils' IAT scores by 0.19-0.2 standard deviation units. The effects translate into females' associating more strongly science with boys and, similarly, girls with humanities disciplines. This finding establishes a new aspect of how educators' stereotypes are transmitted, shaping students' preferences and perceptions of domain-specific adequacy during adolescence.

While the implicit gender attitudes among teachers span substantial changes in girls' implicit gender attitudes, I do not find evidence that assessment-based measures of gender stereotypes do so, at least in this matched sample. The estimates in Column (3) indicate that assessment-based stereotypes are unlikely to shape the student's math-science gender stereotypes. Nevertheless, they have a direct impact on students through channels such as grading (see Appendix Table C.14).

---

[47] The selection criteria for determining the sample of schools that were part of the in-person sessions are the following: (i) the school is in the Lima metropolis, (ii) the school principal provides the authorization for the in-person session, and (iii) the school has verified well-functioning computer labs with internet connection. The educators hired for the data collection either had a bachelor's degree or were senior college students majoring in education or psychology. The 90-minute data-collection sessions took place during classroom time.

[48] The variable $IAT_i$ has been standardized to have mean of zero and standard deviation of one.

[49] This analysis uses different matched teacher-student samples. Columns (1) and (2) are a matched sample of teachers and students who participated in the survey. Column (3) in Table 13 uses a matched administrative dataset of teachers according to enrollment and personnel records matched to students who participated in the survey. In both samples, students correspond to a younger cohort with projected graduation between 2020 and 2025.

# 8 Consequences of exposure to stereotyped assessments in other subjects

This section describes the consequences of educators' stereotyped assessments when exerted in language arts and science classes. To begin, language arts teachers' stereotypical grading has statistically significant detrimental effects on girls' academic progress across all high school grades examined. Columns (1) to (3) of Appendix Table C.10 show that female students exposed to language arts teachers who hold one standard deviation more stereotypical grading practices against girls have timely high school completion rates that are 0.04–0.01 percentage points lower. This pattern is reprised in Column (4), which indicates a differential effect of a similar order to the one reported for mathematics teachers on high school completion after students' projected graduation year. A similarity between mathematics teachers and language arts teachers is shown in the second row of Appendix Table C.10, in which the main effects of gender stereotypical assessments are reported. As with math, the exposure to more stereotyped instructors in language arts brings a net benefit for male students according to the stacked grades estimate displayed in Column (4). However, these positive effects continue to be more persistent than in the case of mathematics teachers, extending beyond the eighth grade and into the ninth grade. Columns (5) to (8) of Appendix Table C.10, which pertain to the likelihood of ever graduating from high school, suggest a similar conclusion.

Another source of more negative consequences for academic progress among girls compared to boys concerns their science teachers. Appendix Table C.11 shows that only the interaction effects of teachers' stereotypical assessments on girls are statistically significant, while the main exposure effects on stereotyped evaluations cease to be significant. Importantly, as shown in Columns (4) and (8) these differential effects are of a lower order—around half the magnitudes found among mathematics and language arts teachers.

Moving on to college-attendance outcomes, the findings I report in Appendix Table C.12 indicate that language arts instructors, unlike math instructors, do not directly affect boys' or girls' college application, admission, or enrollment outcomes. The same pattern regarding science-teacher exposure is presented in Appendix Table C.13. In addition, I consider the recurrent but transitory effects of language arts teachers' stereotyped assessments on girls' labor market outcomes. A conservative interpretation of the results reported in Appendix Tables C.16 to C.20 is that the differential effects of gender stereotypes only appear in formal sector employment, paid hours worked, and monthly earnings for up to two years. These three results are largely repeated in Appendix Tables C.22 to C.26, in which it can be seen that persistence over time is not a factor in exposure to language arts teachers, whose stereotypical grading affect students up to two years after graduating high school. Similarly, the effects of science teachers' stereotypical assessments only last till one year after graduation.

The lower intensity and shorter duration of exposure to stereotyped grading practices among language arts and science teachers in comparison to mathematics educators highlights the well-documented significance of mathematics for academic achievement and advancement.

# 9 Conclusions

I estimated the long-term effects of exposure to stereotypical assessment teachers for 1,7 million students from five graduating cohorts of Peruvian public schools. I followed these students for nearly a decade, beginning when they were 13 years old and in eighth grade and continuing until they were 23 years old and studying in higher education institutions or working. I employed a comprehensive set of administrative and survey data to characterize the distribution of teachers' stereotypical assessments of mathematical and language abilities. My primary measurement of math-science stereotypical assessments is based on the differences between teacher-assigned and centrally-assigned scores for male and female students, using a sample of approximately 60,000 teachers. This is the first study to document that educators' implicit, and to some extent unconscious, gender attitudes are reflected in their *observed* behavior when evaluating their students' performance in mathematics, language arts, and science. In addition, an examination of these two measures enables me to conclude that assessment stereotypes have a strong correlation with implicit attitudes as measured by survey data on a widely administered psychological test.

The central contribution of this study is its estimate that exposure to stereotyped teacher assessments exacerbates gender gaps in earnings and formal employment between ages 18–23 for up to five years after high school graduation. My rich set of controls allowed me to address potential sources of student-to-teacher sorting. I emphasized that educators' stereotypical grading practices in students' assessment are a previously undocumented source of gender gaps in earnings, indicating educators' stereotypes' role in fostering the adverse conditions women face in labor markets.

Another key contribution is my analysis of the mechanisms that allow the effects of gender stereotyped assessments to persist in students' postsecondary trajectories. Teachers' distinct value-added toward female and male students is a potential mechanism, as it provides unabridged educational inputs that, according to the literature, have persistent effects on earnings. I considered the argument that teachers' signaling regarding students' performance can reflect their implicit gender stereotypes regarding students' ability and potentially distort students' beliefs regarding boys' versus girls' sufficiency in science versus humanities. In other words, I examined whether the IAT's measurement of students' implicit stereotypes had changed due to stereotypical grading practices. I cannot conclude that the gender stereotypes expressed in the evaluations of educators are internalized by their students based on the IAT scores of their students.

Given the pervasiveness of teachers' gender stereotypes and their adverse effects on students, an urgent task is to design policy interventions to reduce the prevalence of such stereotypes in educational settings. It also remains to be studied whether students can be trained to recognize teachers' stereotypical assessments and prevent the stereotypes from affecting their human capital decisions.

Figure 1: Map of teachers' survey completion

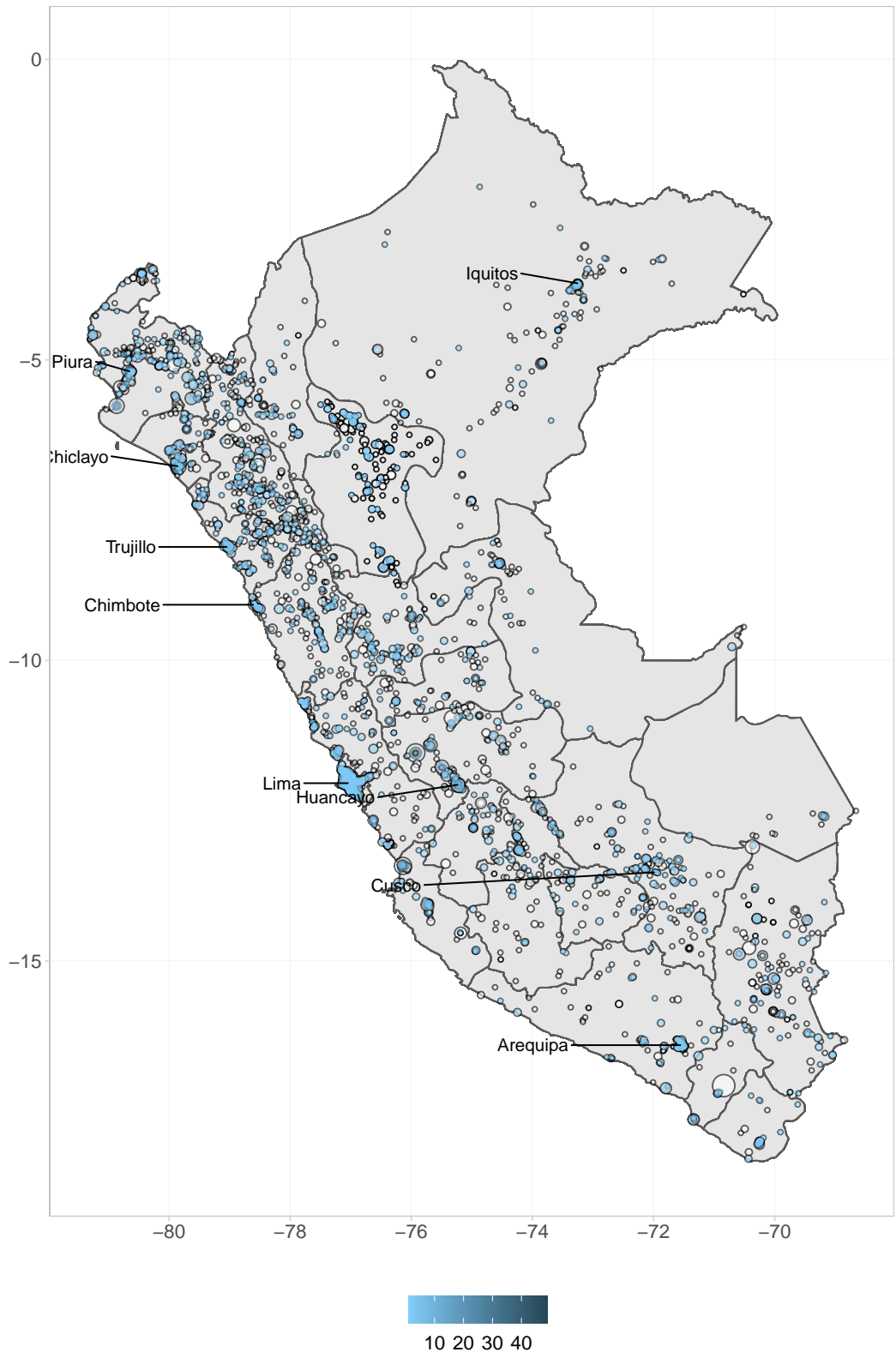

*Notes:* The map plots the number of teachers who registered to complete the survey and those who completed it binned at the school level. The binning uses the school names teachers are employed in as of 2019–20. White circles represent teachers that signed up for the survey but did not complete it, while the blue circles represent teachers who signed up and completed the survey. Teachers in 2,115 schools signed up, and teachers in 1,673 schools completed the survey. The teachers who completed the survey are distributed across schools as follows: 624 schools have one survey-taker teacher (46.9%), 687 (51.7%) have 2–10 survey-taker teachers and 17 schools have between 11 and 49 survey-taker teachers (1.3%).

Figure 2: Data-collection timeline

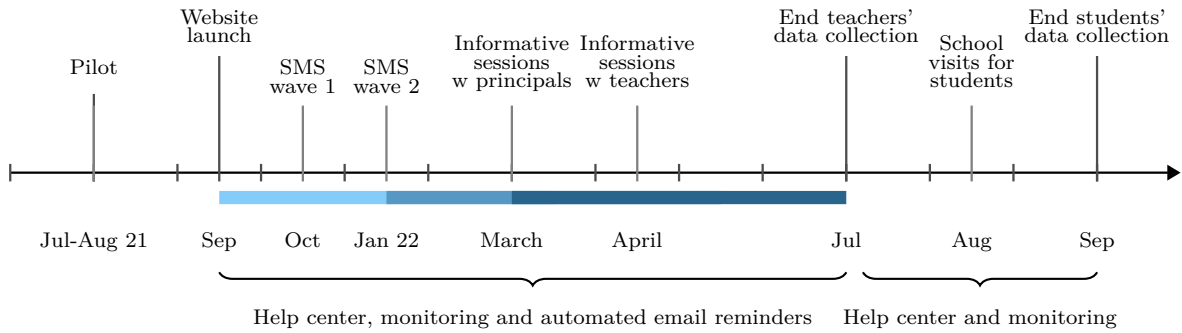

*Notes:* The graph summarizes the data-collection phases that started in September 2021, ended in July 2022 for teachers, and ended in August 2022 for students. The data collection encompassed the last quarter of the school year 2021 and the first three quarters of 2022. The color bar indicates the last quarter of the 2021 school year (the summer-vacation period) and the first two quarters of the 2022 school year, respectively. Before the data collection, instruments and website functionality were tested in a pilot involving around 30 teachers that participated in individual interview sessions. The website was launched in September 2022, starting the data-collection period for teachers. Sampled participants were invited to participate in the website through two waves of SMS. Once teachers registered on the website, they were sent up to four automated email reminders to complete the partially completed activities (that is, the Implicit Association Test [IAT] and the survey). At the start of the 2022 school year, informational meetings were held online for sampled teachers and their principals. Student data collection started in August 2022 and ended in September 2022. Enumerators visited schools in the metropolis of Lima, guiding students to complete the IAT and survey during the homeroom teachers' tutoring hours. Throughout the data-collection and website-functioning period, a help center was available through email, phone, and WhatsApp to aid teachers and students in the remote data collection.

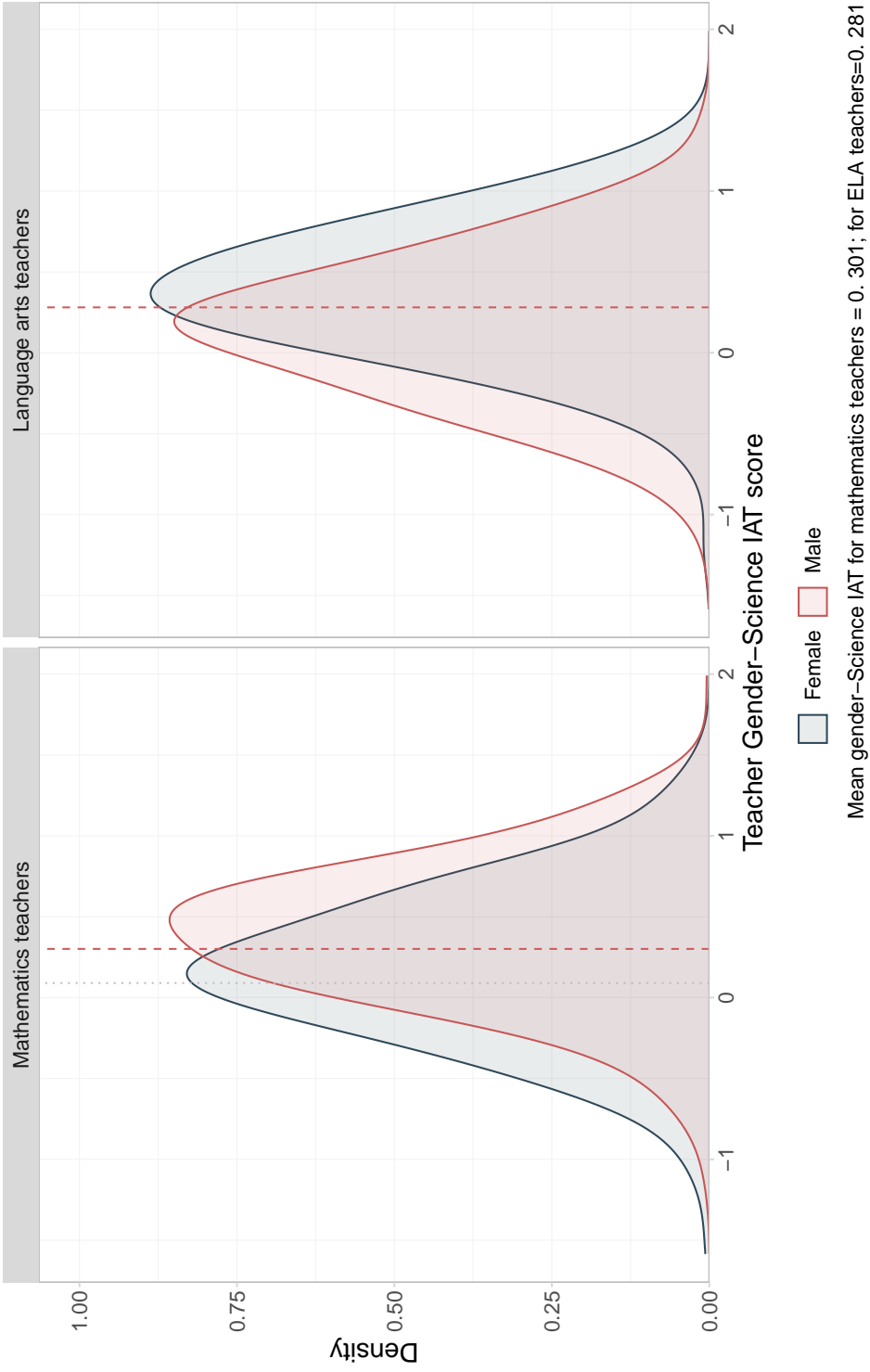

Figure 3: Distribution of teachers' gender-science IAT scores

*Notes:* This figure presents the density of IAT scores for 2,052 (Math= 1,102; ELA=950) teachers successfully matched with administrative records, with an available measure of teacher-level estimated grading gender differences. According to Nosek et al. (2007), IAT scores below -.15 are *in favor of girls*, between -.15 and .15 indicate *little to no stereotypes against girls*, .15 to .35 indicate *slight stereotypes against girls*, and scores above .35 indicate *moderate to severe stereotypes against girls*.

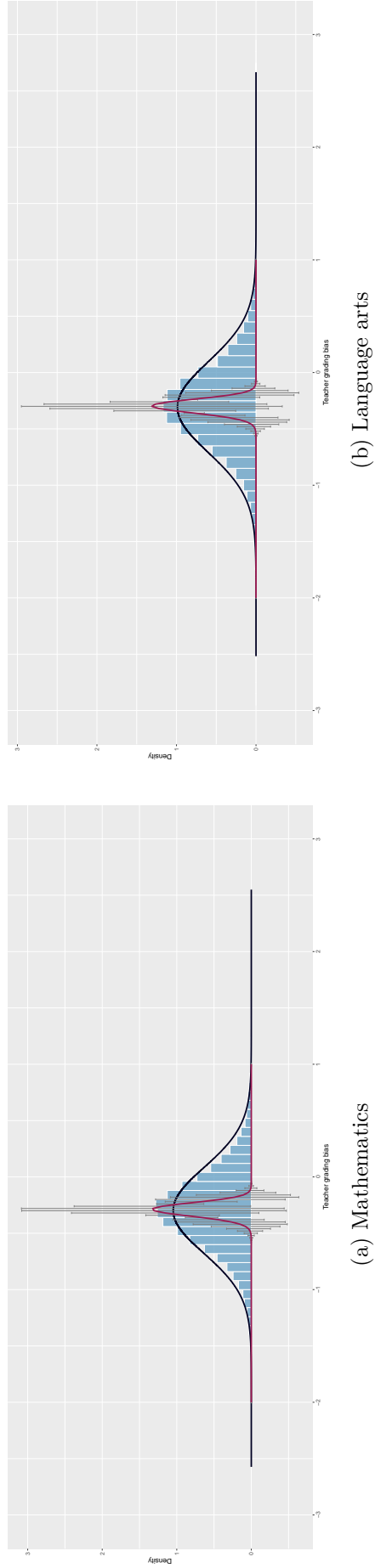

Figure 4: Distribution of posterior means of teacher-level stereotyped grading under EB deconvolution methods

(a) Mathematics  (b) Language arts

*Notes*: The figure presents the EB deconvolution estimates of the prior density of teacher-level stereotyped grading and a benchmark Gaussian density. The histogram shows the distribution of estimated teacher-level stereotyped assessment. The red line density is the prior density of underlying population parameters estimated by the EB deconvolution method proposed in Efron (2016). This density was computed numerically, maximizing the respective (penalized) log-likelihood function using the deconvolveR package developed by Narasimhan and Efron (2020). The complexity penalty parameter was chosen for the deconvolved density to match the mean and bias-corrected variance of the estimated teacher-level stereotyped grading. The parameter support was set to be a finite discrete of points between -2 and 1 for mathematics and language arts. The grey bars correspond to ± one estimated standard error. The black line density corresponds to the Gaussian density with the mean and variance of the posterior deconvolved distribution.

Figure 5: Effects on college application per grade

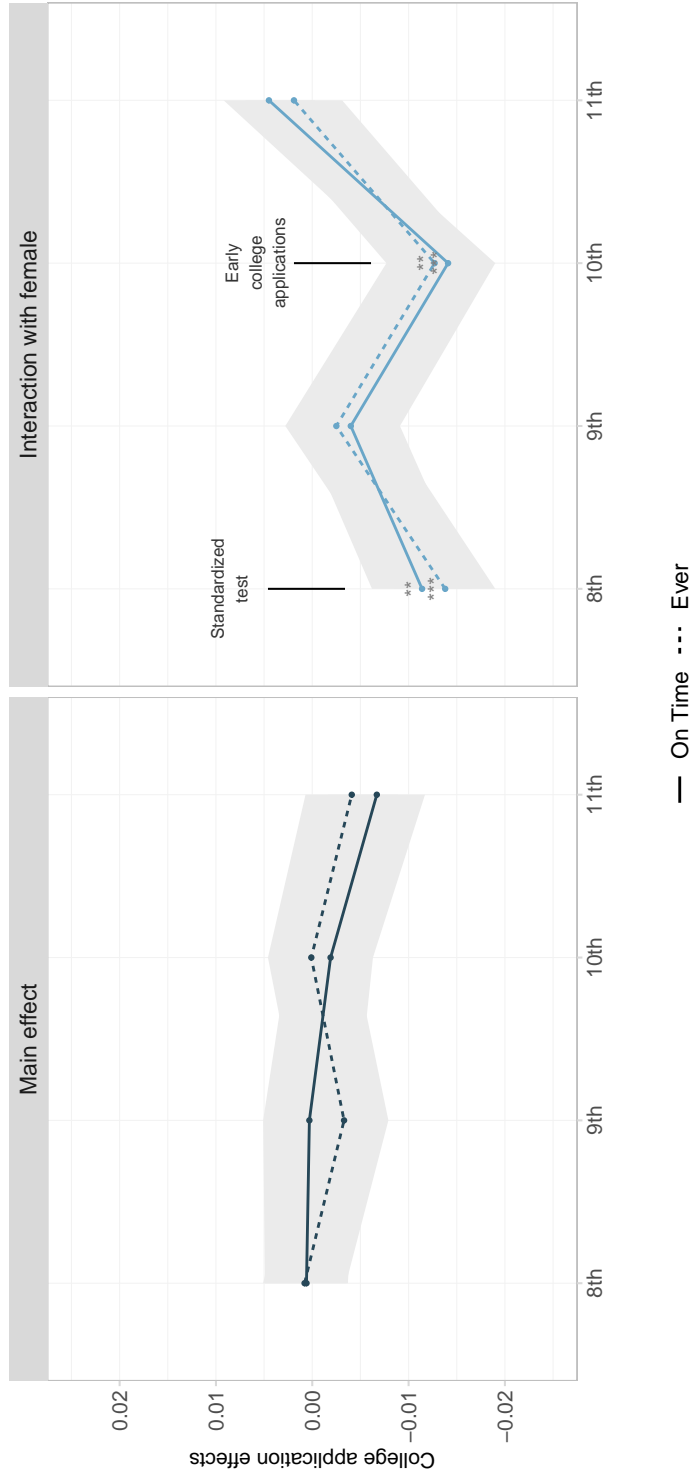

*Notes*: This figure shows the estimated effects of teacher' stereotyped assessments on the likelihood of applying to college on time and ever according to the estimating Equation (12). Panel A reports point estimates of the main effects of exposure to a one-standard-deviation more stereotyped teacher against girls. Panel B reports the interaction effects of exposure with students being female. Clustered standard errors at the school level interacted are shown in shaded regions.

Figure 6: Effects on formal sector employment, ages 18–23

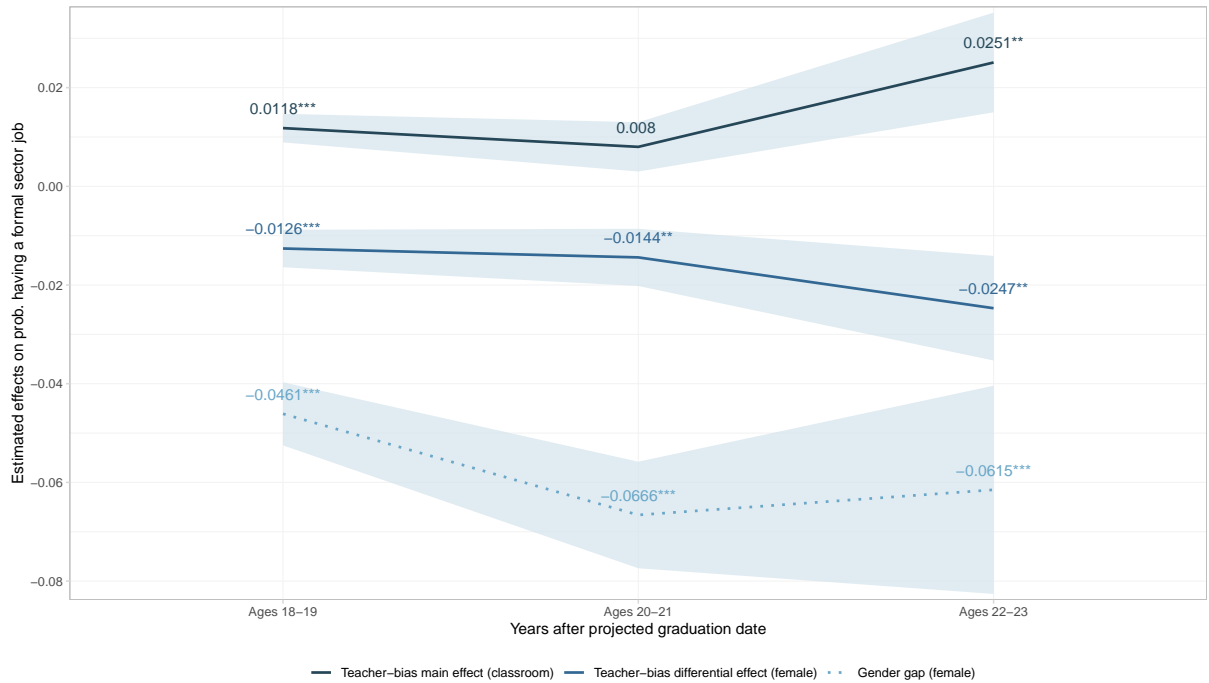

*Notes:* This figure shows the estimated effects of teacher' stereotyped assessments on the likelihood of having a formal sector job at ages 18–23 after graduating from high school. The point estimates of the main effect of exposure to more biased teachers are shown in the top solid line. The second solid line shows point estimates of the differential effects on women for being exposed to a more stereotypical grading teacher. The dotted line shows the effects on women on the likelihood of employment (that is, the gender gap). I include controls according to estimating Equation (12). Clustered standard errors at the school location level, interacting with student gender, are shown in shaded regions.

Figure 7: Effects of teachers' stereotypical assessments on probability of students' earnings being above a percentile after graduating high school, ages 18–19

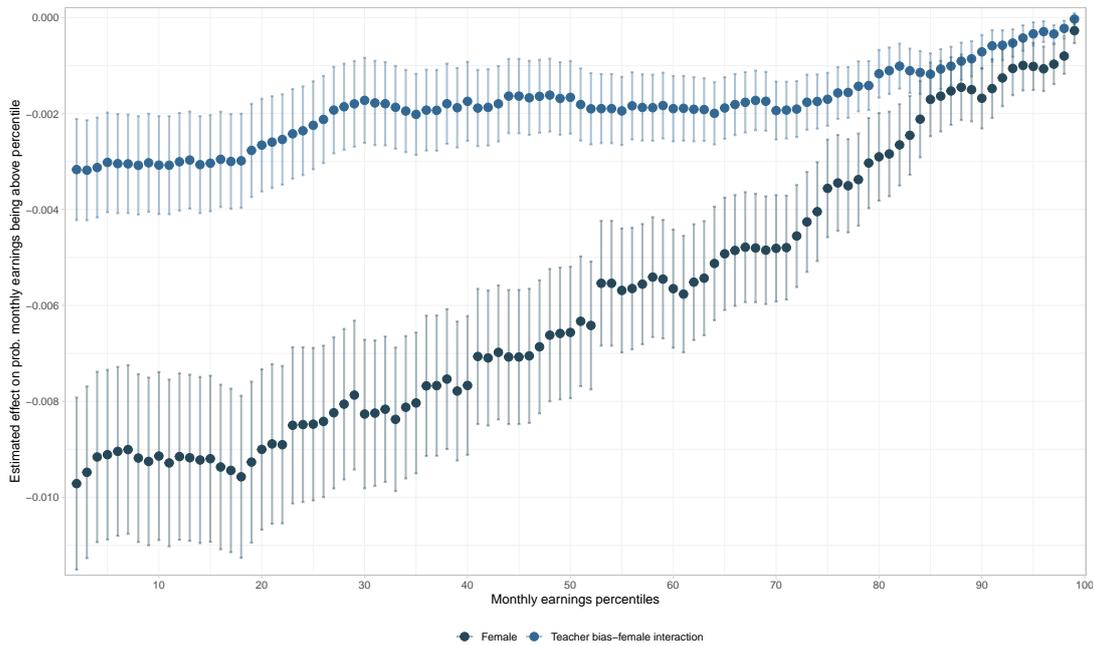

(a) Interaction effects and effects on gender-gap of teacher-level stereotyped assessments

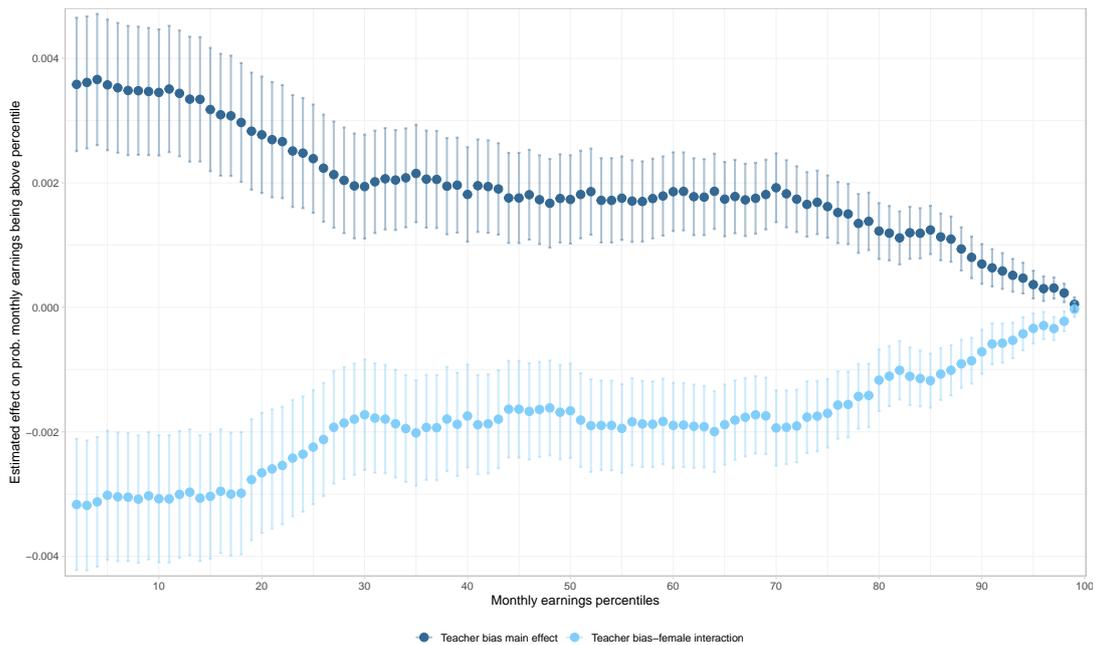

(b) Interaction and main teacher-level stereotyped assessments effects

*Notes:* This graph shows the estimated effects on the probability of monthly earnings of student $i$ being above a given percentile $p$ where $p \in (0, 100)$. The estimates are according to Equation (12), and the percentiles have been computed using the population of workers from the base sample who have between 16 and 25 years old.

Table 1: Descriptive statistics for eighth-grade students with teacher-assessed and blindly graded test scores

|  | Female (1) | Male (2) | Total (3) |
|---|---|---|---|
| *A. Demographic characteristics* | | | |
| Age, years | 12.750 | 12.873 | 12.812 |
| Spanish | 0.589 | 0.586 | 0.587 |
| Indigenous language (Quechua) | 0.076 | 0.079 | 0.078 |
| Other | 0.017 | 0.018 | 0.017 |
| Missing language | 0.318 | 0.317 | 0.318 |
| Family receives CCT | 0.231 | 0.246 | 0.239 |
| Mother main caregiver | 0.688 | 0.672 | 0.680 |
| Father main caregiver | 0.291 | 0.308 | 0.300 |
| Other main caregiver | 0.021 | 0.020 | 0.020 |
| *B. School attributes* | | | |
| School in Lima | 0.223 | 0.211 | 0.217 |
| School in urban area | 0.869 | 0.857 | 0.863 |
| *C. Students' educational aspirations* | | | |
| HS | 0.058 | 0.075 | 0.067 |
| Technical | 0.059 | 0.109 | 0.084 |
| College | 0.287 | 0.292 | 0.290 |
| Graduate | 0.259 | 0.182 | 0.220 |
| Missing | 0.337 | 0.342 | 0.339 |
| Num. of students | 690,449 | 696,800 | 1,387,249 |
| Num. of schools |  |  | 9,358 |
| Num. of teachers |  |  | 61,419 |
| Observations |  |  | 3,513,562 |

*Notes:* This table reports summary statistics on eighth-grade students who attended schools during 2015, 2016, 2018, and 2019. Students in the sample have available information on the standardized examination and teacher-assigned test scores in mathematics, language arts, and science. In addition, pupils in the sample have been successfully matched with their respective subject teachers. The bottom rows report the number of schools the students in the sample attend and the number of teachers matched to these students. Observations are at the student-subject level.

Table 2: Descriptive statistics for 8th- to 11th-grade of high school teacher-student matched sample

|  | Base sample | | | Full regression sample | | |
|---|---|---|---|---|---|---|
|  | Male (1) | Female (2) | Total (3) | Male (4) | Female (5) | Total (6) |
| *A. Demographic characteristics* | | | | | | |
| Spanish | 0.871 | 0.875 | 0.873 | 0.841 | 0.842 | 0.842 |
| Indigenous language (Quechua) | 0.104 | 0.100 | 0.102 | 0.130 | 0.128 | 0.129 |
| Other language | 0.024 | 0.025 | 0.025 | 0.028 | 0.030 | 0.029 |
| Born in Lima | 0.239 | 0.246 | 0.243 | 0.195 | 0.201 | 0.198 |
| Mother is main caregiver | 0.702 | 0.720 | 0.711 | 0.682 | 0.698 | 0.689 |
| *B. Students' senior year* | | | | | | |
| 2015 | 0.062 | 0.066 | 0.064 | 0.046 | 0.049 | 0.047 |
| 2016 | 0.133 | 0.140 | 0.136 | 0.105 | 0.110 | 0.108 |
| 2017 | 0.204 | 0.212 | 0.208 | 0.170 | 0.175 | 0.173 |
| 2018 | 0.224 | 0.227 | 0.226 | 0.254 | 0.261 | 0.258 |
| 2019 | 0.239 | 0.240 | 0.239 | 0.278 | 0.283 | 0.280 |
| Observations | 7,876,733 | 7,508,521 | 15,385,254 | 4,169,532 | 3,816,044 | 7,985,576 |
| Num. of students | 1,083,084 | 1,048,483 | 2,131,560 | 890,078 | 846,049 | 1,736,120 |
| Num. of teachers |  |  | 122,591 |  |  | 53,975 |

*Notes:* This table reports summary statistics on 8th- to 11th-grade students enrolled in school in 2015-2019 with available end-of-grade scores for a previous school year. Columns 1 to 3 describe the *base* sample, which includes all students successfully matched with a teacher who instructs mathematics, language arts, or science. Columns 4 to 6 report statistics on the *full* regression sample, which excludes students in the base sample whose teachers' lack information on their stereotypical assessment measure. The observations are at the student-grade-subject level.

Table 3: Descriptive statistics for students after high school graduation: labor market characteristics

|  | Base sample | | | Workers in regression sample | |
|---|---|---|---|---|---|
|  | Female (1) | Male (2) | Non reported (3) | Female (4) | Male (5) |
| *A. Average earnings and hours, 2015-2020* | | | | | |
| Monthly earnings (2010 USD) | 218.113 | 254.862 | 321.517 | 218.702 | 255.303 |
| Paid work hours per month | 40.940 | 46.902 | 65.518 | 41.036 | 46.791 |
| Hourly wage (2010 USD) | 2.759 | 2.979 | 3.421 | 2.765 | 2.986 |
| *B. Worker characteristics* | | | | | |
| Worker age, years | 19.202 | 19.244 | 19.437 | 19.209 | 19.246 |
| Special education | 0.000 | 0.000 | 0.000 | 0.001 | 0.001 |
| Less than HS | 0.046 | 0.064 | 0.163 | 0.045 | 0.063 |
| HS graduate | 0.621 | 0.691 | 0.779 | 0.616 | 0.687 |
| Some college or technical | 0.195 | 0.118 | 0.039 | 0.194 | 0.118 |
| Technical degree | 0.051 | 0.046 | 0.019 | 0.056 | 0.049 |
| No information | 0.088 | 0.082 | 0.000 | 0.085 | 0.079 |
| *C. NAICS industry* | | | | | |
| Firm size (num. workers per firm) | 3,850.377 | 3,185.798 | 2,334.833 | 3,809.744 | 3,173.817 |
| 11 agriculture, forestry, fishing | 0.218 | 0.224 | 0.270 | 0.216 | 0.223 |
| 21-23 mining, utilities, construction | 0.017 | 0.082 | 0.073 | 0.017 | 0.082 |
| 31-33 manufacturing | 0.180 | 0.179 | 0.253 | 0.178 | 0.178 |
| 42-49 trade, transportation | 0.222 | 0.213 | 0.150 | 0.222 | 0.213 |
| 51-59 information, finance, prof. serv. | 0.183 | 0.167 | 0.124 | 0.184 | 0.167 |
| 61-62 educational and health care serv. | 0.019 | 0.008 | 0.006 | 0.020 | 0.009 |
| 71-72 arts, recreation, hospitality serv. | 0.099 | 0.074 | 0.058 | 0.100 | 0.074 |
| 81 other services | 0.057 | 0.046 | 0.060 | 0.057 | 0.046 |
| 92-99 public admin., unclas. | 0.005 | 0.007 | 0.004 | 0.005 | 0.007 |
| Number of employees | 176,963 | 281,511 | 2,014,575 | 187,919 | 297,087 |
| Number of firms | 30,214 | 45,798 | 190,055 | 31,990 | 47,667 |
| Observations | 176,963 | 281,511 | 2,014,575 | 187,919 | 297,087 |

*Notes:* This table shows descriptive statistics for comparing characteristics of recently graduated workers and students constituting the high school estimation sample with available matched employer-employee records. Students appearing in these records were reported by an employer to have a formal sector job corresponding to January 2015 and December 2020. Columns 1 to 3 show means for workers in the base sample, comprising workers between 16 and 25 years old. This age range retrieves comparable work experience to cohorts of recently graduated high school students in the regression sample. Workers in the base sample also comply with all exclusion filters described in the Online Appendix Section A.2.1. Columns 4 and 5 show means for students with projected graduation years 2015-20 who (i) are part of the estimation sample and thus have been matched with public high school student records, and (ii) appear as being employed for at least one month over the period 2015-2020 in administrative employment records. Panel A reports the means of monthly earnings, paid working hours, and hourly wages that have been previously processed to reflect the compensations from the work contact with the *dominant annual employer* as defined in Online Appendix Section A.2.1. Panels B and C report workers' characteristics and occupation categories as recorded by the most recent work contact with a *dominant annual employer*.

Table 4: Descriptive statistics for surveyed teachers with implicit-stereotypes measure

|  | Mathematics (1) | Language arts (2) | Total (3) |
|---|---|---|---|
| *A. Demographic characteristics* | | | |
| Female | 0.454 | 0.543 | 0.495 |
| Age, years | 42.977 | 41.540 | 42.312 |
| Mixed | 0.650 | 0.660 | 0.654 |
| Indigenous language (Quechua) | 0.227 | 0.204 | 0.216 |
| White | 0.022 | 0.019 | 0.020 |
| Asian | 0.003 | 0.002 | 0.002 |
| Afroperuvian | 0.031 | 0.025 | 0.028 |
| Has at least one daughter | 0.532 | 0.482 | 0.509 |
| Has no daughters | 0.257 | 0.245 | 0.251 |
| Has neither daughter nor sons | 0.211 | 0.273 | 0.239 |
| *B. Teaching appointment* | | | |
| School in Lima | 0.240 | 0.253 | 0.246 |
| Teaching hours per week | 28.438 | 27.811 | 28.148 |
| Number of students per class | 24.222 | 24.983 | 24.574 |
| Classoom teacher | 0.606 | 0.581 | 0.595 |
| First grade | 0.532 | 0.563 | 0.546 |
| Second grade | 0.562 | 0.590 | 0.575 |
| Third grade | 0.531 | 0.649 | 0.585 |
| Fourth grade | 0.554 | 0.627 | 0.587 |
| Fifth grade | 0.552 | 0.586 | 0.568 |
| *C. Job characteristics* | | | |
| Tenure | 0.430 | 0.401 | 0.417 |
| Contract | 0.518 | 0.554 | 0.535 |
| No contract | 0.018 | 0.019 | 0.019 |
| Other contract | 0.034 | 0.026 | 0.030 |
| Public sch experience, years | 12.635 | 11.382 | 12.055 |
| Private sch experience, years | 2.647 | 3.030 | 2.824 |
| College major, education | 0.851 | 0.914 | 0.880 |
| College major, other | 0.149 | 0.086 | 0.120 |
| College | 0.644 | 0.655 | 0.649 |
| Technical institution | 0.447 | 0.419 | 0.434 |
| Other institution | 0.015 | 0.013 | 0.014 |
| *D. Discrimination and IAT* | | | |
| Experienced discrimination | 0.164 | 0.185 | 0.173 |
| Witnessed discrimination | 0.175 | 0.190 | 0.182 |
| Gender-science stereotype (IAT score) | 0.301 | 0.281 | 0.292 |
| Gender-career stereotype (IAT score) | 0.265 | 0.270 | 0.267 |
| Observations | 1,102 | 950 | 2,052 |

*Notes:* This table reports descriptive statistics for the analysis sample of high school teachers (grades 7 to 11) who were surveyed and matched with their stereotypical assessment estimates. Teachers' age and teaching experiences were reported between September 2021 and July 2020. Teaching appointment and job characteristics reported correspond to the years 2021 and 2022. Experience and witnessing discrimination is a self-reported measure of discriminatory behaviors by colleagues or principals during the school year 2021-2022 on the basis of race, immigrant status, socioeconomic level, gender, sexual identity, and religion. Raw (non-standardized) IAT scores are reported. Teachers have *strong stereotypes* if their raw IAT score is above .65, *moderate to severe stereotypes* when their score is between .65 and .35, *slight stereotypes* if their score is between .15 and .35, *little to no stereotypes* if their score is between -.15 and .15, and *preference for girls* if the score is below -.15.

Table 5: Variation in estimates of teacher-level stereotypical assessments

|  | Teachers' subjects | | |
|---|---|---|---|
|  | Mathematics (1) | Language arts (2) | Science (3) |
| Unadjusted mean | -0.2978 | -0.3069 | -0.4119 |
| Bias corrected standard deviation |  |  |  |
|    Unweighted | 0.0612 | 0.0706 | 0.0880 |
|    Student-weighted | 0.0973 | 0.1106 | 0.1576 |
| Num. of teachers | 21,094 | 21,078 | 13,482 |

*Notes:* This table presents the estimated means and standard deviations of the estimated stereotyped teacher assessments, $\hat{\theta}_j$, calculated from estimates of Equations (1) and (2). The first row reports the mean, $\hat{\mu}$, of the unadjusted $\hat{\theta}_j$. The second and third rows display bias-corrected variance estimates that use the associated standard errors $s_j$ to correct for bias due to sampling error in $\hat{\theta}_j$. The second row reports an *unweighted* bias-corrected variance, $\hat{\sigma}_U$, and the third row presents the bias-corrected *weighted* variance, $\hat{\sigma}_W$, where the sampling error was computed with student weights are $w_j = \mathcal{N}_j/\mathcal{N}$ with $\mathcal{N}(j)$ equivalent to the number of students assigned to teacher $j$ and $\mathcal{N} = \sum_j \mathcal{N}(j)$. Observations are at the teacher level. Standard errors were computed using clusters at the student level.

Table 6: Relationship between gender differences in assessment and mathematics teachers' characteristics

|  | (1) | (2) | (3) |
|---|---|---|---|
| *A. Demographic characteristics* | | | |
| Female | 0.3269*** | 0.3286*** | 0.4455*** |
|  | (0.043) | (0.043) | (0.0789) |
| Age, older than median | -0.2312*** | -0.2523*** | -0.086 |
|  | (0.0497) | (0.0518) | (0.093) |
| Higher ed., university | -0.1503** | -0.0835 | -0.0808 |
|  | (0.0652) | (0.0669) | (0.0813) |
| *B. Teaching experience in private schools, years* | | | |
| Less than 2 yrs |  | -0.3688*** | -0.4137*** |
|  |  | (0.0928) | (0.1099) |
| 2-5 yrs |  | -0.1749** | -0.2173** |
|  |  | (0.0816) | (0.0967) |
| 6-10 yrs |  | -0.2806** | -0.3658*** |
|  |  | (0.1153) | (0.1415) |
| More than 10 yrs |  | -0.4032** | -0.5239** |
|  |  | (0.1681) | (0.2034) |
| *C. Teaching experience in public schools, years* | | | |
| Less than 2 yrs |  | -0.2994 | -0.1336 |
|  |  | (0.2584) | (0.5585) |
| 2-5 yrs |  | -0.4028* | -0.2201 |
|  |  | (0.2326) | (0.5278) |
| 6-10 yrs |  | -0.191 | -0.0855 |
|  |  | (0.2326) | (0.5279) |
| More than 10 yrs |  | -0.2262 | -0.1952 |
|  |  | (0.2377) | (0.531) |
| *D. Teacher evaluation performance* | | | |
| Skill and knowledge test, z score |  |  | -0.1473** |
|  |  |  | (0.0586) |
| Passed test |  |  | -0.0616 |
|  |  |  | (0.1455) |
| R-squared | 0.0057 | 0.0058 | 0.0124 |
| Observations | 20,411 | 20,411 | 7,161 |

*Notes:* This table reports the estimated relationship between mathematics teachers' stereotypical assessment estimates and a set of covariates. The reported coefficients were estimated according to Equation 4, and the outcome variable was divided by the bias-corrected student-weighted standard deviation. The median age of teachers is 45 years old in the sample as of 2019. The base category for indicators of higher education is *Technical Institute*. *Teaching experience in public schools* and *Teaching experience in private schools* are determined using 2019 personnel records from the Ministry of Education. The base category for the experience variables is *No experience*. Teacher-evaluation performance covariates are drawn from the National Teacher Evaluations from 2015, 2017, 2018, and 2019. *Skills and knowledge test z score* is the standardized score of the national-stage evaluation of knowledge and teaching skills designed to test teachers applying to permanent public-school teaching positions. *Passing status* indicates that the the teacher has approved the national-stage exam in any of the preceding years. Observations are at the teacher level. I include school-fixed effects and missing-value dummies where the information is unavailable. Standard errors are clustered at the school level.

Table 7: Relationship between gender differences in assessment and implicit gender stereotypes

|  | Mathematics teachers | Language arts teachers |
|---|---|---|
|  | Gender-science (1) | Gender-science (2) |
|  | A. No controls | |
| IAT score | 0.1682*** | -0.4943*** |
|  | (0.0489) | (0.045) |
| R-squared | 0.035 | 0.0882 |
| Observations | 1,102 | 950 |
|  | B. With covariates | |
| IAT score | 0.1279** | -0.5138*** |
|  | (0.0501) | (0.045) |
| R-squared | 0.1127 | 0.1272 |
| Observations | 1,102 | 950 |

*Notes:* This table reports regression estimates of implicit stereotypes, measured with Implicit Association Test (IAT) scores, on the stereotyped assessment of mathematics and language arts teachers, according to Equation (5). The IAT score has been standardized to have a mean of 0 and a standard deviation of 1. Column 1 shows the effect of a one standard deviation increase in math teachers' math-science stereotypes (as measured by the IAT score) on their gender gap in mathematics grading. Column 2 reports similar regression effects of increasing by one standard deviation language arts teachers' implicit gender math-science stereotypes against girls on their grading gender gap in language arts. In Column 1, the dependent variable was divided by the bias-corrected standard deviation of $\hat{\theta}_j$ for math teachers, while a similar procedure was used for language arts teachers in Column 2. Therefore, the reported coefficients are expressed in standard-deviation units. The first panel reports estimates including only school-location fixed effects. The second panel adds control for teachers' gender, childbearing status (i.e., daughters, sons, or both), indicators of the decades of teachers' date of birth, race, the order of the associations in the IAT (i.e., order compatible or order incompatible associations appearing first), and the number of previous IATs the teacher has taken. Observations are at the teacher level, and clustered robust standard errors by school location are reported in parentheses.

Table 8: High school graduation effects of exposure to mathematics-teacher stereotyped assessments

| | Graduated on time | | | | Graduated ever | | | |
|---|---|---|---|---|---|---|---|---|
| | 8th grade (1) | 9th grade (2) | 10th grade (3) | Stacked grades (4) | 8th grade (5) | 9th grade (6) | 10th grade (7) | Stacked grades (8) |
| $\hat{\theta}^*_{j,-i} \cdot$ Female | -0.0364*** | -0.0173*** | -0.0147*** | -0.0151*** | -0.0387*** | -0.0135*** | -0.0136*** | -0.0133*** |
| | (0.0054) | (0.0045) | (0.0037) | (0.0023) | (0.0053) | (0.0043) | (0.0034) | (0.0022) |
| $\hat{\theta}^*_{j,-i}$ | 0.0208*** | 0.0029 | 0.002 | 0.0051* | 0.0227*** | 0.0028 | 0.0021 | 0.006** |
| | (0.0059) | (0.0061) | (0.0051) | (0.0027) | (0.0057) | (0.006) | (0.005) | (0.0028) |
| Female | -0.0054 | -0.0226** | -0.0097 | -0.0089** | -0.0197** | -0.0376*** | -0.0204*** | -0.0191*** |
| | (0.0103) | (0.0088) | (0.0069) | (0.0041) | (0.01) | (0.0086) | (0.0065) | (0.0041) |
| Baseline controls | × | × | × | × | × | × | × | × |
| Lagged scores | × | × | × | × | × | × | × | × |
| $\bar{Y}$ female | 0.7513 | 0.7883 | 0.8572 | 0.8239 | 0.782 | 0.8249 | 0.8933 | 0.8544 |
| $\bar{Y}$ male | 0.6833 | 0.7327 | 0.8105 | 0.7741 | 0.7434 | 0.8019 | 0.8784 | 0.8318 |
| R-squared | 0.3271 | 0.364 | 0.4275 | 0.4189 | 0.2674 | 0.2871 | 0.2897 | 0.2831 |
| Observations | 629,489 | 667,250 | 765,620 | 2,905,775 | 629,489 | 667,250 | 765,620 | 2,905,775 |

*Notes*: This table reports estimates of the effects of one-grade exposure to a mathematics teacher's stereotypical assessment practices on high school graduation on time and ever. The estimations use the full regression sample of students between grades 8 and 10 with projected graduation years between 2015 and 2019. The teacher-level stereotyped grading in assessment is a leave-one-year-out estimate assigned to each student cohort based on their projected graduation year. The outcome variable in Columns 1 to 4 is an indicator variable taking value one when the student graduated in the school year corresponding to their projected high school graduation date. According to enrollment records, the projected graduation year was computed based on the earliest year the student was enrolled in high school. Columns 5 to 8 have as outcome variables an indicator for graduation at any time. The window of time over which graduating from high school *ever* is defined varies cohort-to-cohort, ranging from one for the youngest cohort (class of 2019) to four calendar years after projected graduation for the oldest cohort (class of 2015). Covariates include baseline student-level characteristics and teacher attributes interacted with student-gender, classroom and school-level controls, and quadratic polynomials of lagged mathematics and language arts scores. The observation unit in all-grades estimation (Columns 4 and 8) is the student; this specification includes cohort, year, and school fixed effects and clustered standard errors at the school level. The observation unit in grade-specific estimations (Columns 1 to 3 and Columns 5 to 7) is the student; this specification uses a sample of students stacked across grades and includes grade, cohort, year, and school fixed effects, as well as two-way clustered standard errors by student and school.

Table 9: College-attendance effects of exposure to mathematics-teacher stereotypical assessment

| | College application | | College admission | | College enrollment | |
|---|---|---|---|---|---|---|
| | On time (1) | Ever (2) | On time (3) | Ever (4) | On time (5) | Ever (6) |
| $\hat{\theta}^*_{j,-i} \cdot$ Female | -0.0066** | -0.0073** | -0.0011 | -0.001 | -0.0006 | -0.0011 |
| | (0.0032) | (0.0032) | (0.0026) | (0.0027) | (0.0026) | (0.0026) |
| $\hat{\theta}^*_{j,-i}$ | 0.0007 | 0.0015 | -0.0028 | -0.0028 | -0.0031 | -0.0032 |
| | (0.0029) | (0.0027) | (0.0022) | (0.0022) | (0.0022) | (0.0022) |
| Female | 0.0161*** | 0.0076 | 0.0063 | 0.0042 | 0.0056 | 0.0051 |
| | (0.0054) | (0.0056) | (0.0044) | (0.0045) | (0.0042) | (0.0042) |
| Baseline controls | × | × | × | × | × | × |
| Lagged scores | × | × | × | × | × | × |
| $\bar{Y}$ female | 0.2438 | 0.3334 | 0.1987 | 0.2077 | 0.1806 | 0.1843 |
| $\bar{Y}$ male | 0.1944 | 0.281 | 0.1644 | 0.1735 | 0.151 | 0.1538 |
| R-squared | 0.1048 | 0.1323 | 0.0869 | 0.0955 | 0.0901 | 0.0928 |
| Observations | 2,905,775 | 2,905,775 | 2,905,775 | 2,905,775 | 2,905,775 | 2,905,775 |

*Notes*: This table reports estimates of the effects of one-grade exposure to a mathematics teacher's stereotypical assessment practices on college-attendance outcomes. The estimation uses a stacked sample of students between grades 8 and 11 (13 to 16 years old) with projected graduation years between 2015 and 2019 that are part of the full regression sample described in Section 3.2.2. The estimation controls for the set of covariates described in the notes of Table 8, which includes grade, cohort, year, and school fixed effects. Two-way clustered standard errors by student and school are reported in parentheses.

Table 10: Bias-corrected correlation matrix of parameters of teacher value-added and stereotyped teacher assessments

|  | Assessment bias (1) | Teacher VA parameters | |
|---|---|---|---|
|  |  | VA towards females (2) | VA towards males (3) |
| A. Mathematics | | | |
| Assessment bias | 1 | -0.1157 | -0.2143 |
| VA toward females |  | 1 | 0.9884 |
| VA toward males |  |  | 1 |
| Standard deviation | 0.1160 | 0.5758 | 0.5986 |
| B. Language arts | | | |
| Assessment bias | 1 | -0.3642 | -0.6577 |
| VA toward females |  | 1 | 0.8673 |
| VA toward males |  |  | 1 |
| Standard deviation | 0.1240 | 0.1259 | 0.1793 |

*Notes:* This table reports the correlation matrix between teacher-level assessment bias and value-added parameters. Numbers in brackets are standard deviations of the parameters reported in each column. I define the assessment bias $\theta_j = \beta_{2,j} - \alpha_{2,j}$, teachers' value-added for males $\alpha_{1,j} + \alpha_{2,j}$, and value-added for females $\alpha_{1,j}$. The parameters were estimated using a fixed-effects-SURE specification of according to Equation (1) and Equation (2). The variance-covariance matrix of joint parameters was adjusted for sampling error using SURE-calculated sampling covariances. The estimating equations control for quadratic polynomial lagged scores in language and math, student-level, and teacher-level baseline controls, classroom-grade and school-grade means of baseline covariates, and cohort, grade, school, and year fixed effects.

Table 11: Employment in the formal sector: effects of exposure to mathematics teachers' stereotyped assessments

| | Employed in formal sector | | |
|---|---|---|---|
| | Ages 18–19 (1) | Ages 20–21 (2) | Ages 22–23 (3) |
| $\hat{\theta}^*_{j,-i} \cdot$ Female | -0.0126*** | -0.0144** | -0.0247** |
| | (0.0038) | (0.0058) | (0.0106) |
| $\hat{\theta}^*_{j,-i}$ | 0.0118*** | 0.008 | 0.0251** |
| | (0.0029) | (0.005) | (0.0101) |
| Female | -0.0461*** | -0.0666*** | -0.0615*** |
| | (0.0064) | (0.0108) | (0.0211) |
| Baseline controls | × | × | × |
| Lagged scores | × | × | × |
| $\bar{Y}$ female | 0.0694 | 0.0963 | 0.1153 |
| $\bar{Y}$ male | 0.1154 | 0.151 | 0.1743 |
| Observations | 884,231 | 395,940 | 118,550 |

*Notes*: This table reports estimated coefficients of the effects of exposure to more biased teachers during one grade in high school on the probability of having a formal sector job at ages 18 to 23 after high school graduation. The sample includes students in the *full regression sample* with projected graduation between 2015 and 2019. The outcome variable is the likelihood of having a formal sector job with a *dominant annual employer* for at least one month in the year $t$ after graduating from high school when students' modal ages are 18–19, 20–21, and 22–23. See Appendix A for details on this variable calculation. The treatment variable was computed as a leave-one-year-out version of teacher-level grading bias, divided by its respective standard deviation computed at the year-subject level. Covariates include student-level characteristics, teacher characteristics, lagged scores, classroom- and school-grade-level controls, and cohort, grade, and year fixed effects. Additionally, I controlled for school-location fixed effects are interacted with gender. The observation unit is the student grade. Clustered standard errors by the school are reported in parentheses.

Table 12: Paid hours work and monthly earnings: effects of exposure to mathematics teachers' stereotyped assessments

| | Paid monthly work hours | | | Monthly earnings (2010 dollars) | | |
|---|---|---|---|---|---|---|
| | Ages 18–19 (1) | Ages 20–21 (2) | Ages 22–23 (3) | Ages 18–19 (4) | Ages 20–21 (5) | Ages 22–23 (6) |
| $\hat{\theta}^*_{j,-i} \cdot$ Female | -1.0624*** | -0.8031 | -1.6004 | -2.6219*** | -1.304 | -3.0888 |
| | (0.3679) | (0.5758) | (1.0464) | (0.7671) | (1.1849) | (2.2144) |
| $\hat{\theta}^*_{j,-i}$ | 1.0016*** | 0.2338 | 2.7815*** | 2.4911*** | 0.2876 | 5.281*** |
| | (0.3041) | (0.5005) | (0.9513) | (0.6362) | (1.045) | (2.0422) |
| Female | -3.8758*** | -6.1869*** | -5.0676** | -8.2754*** | -11.181*** | -11.0696** |
| | (0.6472) | (1.1269) | (2.2476) | (1.3323) | (2.3933) | (4.6696) |
| Baseline controls | × | × | × | × | × | × |
| Lagged scores | × | × | × | × | × | × |
| $\bar{Y}$ female cond. | 96.6 | 98.3 | 98.5 | 212.7 | 220.7 | 223.7 |
| $\bar{Y}$ male cond. | 103.4 | 103.4 | 102.2 | 247.3 | 251.3 | 251.5 |
| $\bar{Y}$ female uncond. | 5.5 | 7.7 | 8.6 | 9.8 | 14.2 | 16.6 |
| $\bar{Y}$ male uncond. | 9.8 | 12.8 | 13.8 | 18.6 | 24.8 | 27.8 |
| Observations | 884,231 | 395,940 | 118,550 | 884,231 | 395,940 | 118,550 |

*Notes*: This table reports estimated coefficients of the effects of exposure to more biased teachers during one grade in high school on paid monthly work hours and monthly earnings at ages 18 to 23 after high school graduation. The sample includes students in the *full regression sample* with projected graduation between 2015 and 2019. In Columns 1 to 3 the outcome variable is the number of work hours paid per month in the contract with the dominant annual employer, and in Columns 4 to 6 the outcome variable is monthly earnings measured in 2010 USD obtained in the contract with the dominant annual employer. See Appendix A for details on this variable calculation. The treatment variable was computed as a leave-one-year-out version of teacher-level grading bias, divided by its respective standard deviation computed at the year-subject level. Covariates include student-level characteristics, teacher characteristics, lagged scores, classroom- and school-grade-level controls, and cohort, grade, and year fixed effects. Additionally, I controlled for school-location fixed effects interacted with gender. The observation unit is the student grade. Clustered standard errors by the school are reported in parentheses.

Table 13: Mechanisms analysis of internalized mathematics teachers' stereotyped assessments among students

|  | Teachers' implicit gender bias, $IAT_j$ | | Teachers' assessment based gender bias, $\hat{\theta}^*_{j,-i}$ |
|---|---|---|---|
|  | (1) | (2) | (3) |
| $GenderBias_j \cdot$ Female | 0.1963* | 0.2028* | -0.1496 |
|  | (0.0989) | (0.1001) | (0.1898) |
| $GenderBias_j$ | -0.2019* | -0.1888 | 0.1515 |
|  | (0.1) | (0.105) | (0.1737) |
| Female | 0.1411* | 0.1126* | 0.0331 |
|  | (0.0722) | (0.0618) | (0.3567) |
| Mean of outcome female | 0.0985 | 0.0985 | 0.2967 |
| Mean of outcome male | 0.0755 | 0.0755 | 0.3212 |
| Baseline controls | No | Yes | Yes |
| R-squared | 0.0026 | 0.0241 | 0.1187 |
| Observations | 1,052 | 1,052 | 4,640 |

*Notes* This table reports the relationship between a one-standard-deviation increase in teachers' gender bias against girls and their students' Implicit Association Test (IAT) scores. The average IAT among students that took the survey is 0.097 for females and 0.072 for males. The dependent variable (i.e., students' gender-science IAT score) was standardized with a mean of 0 and a standard deviation of 1. Columns (1) and (2) report the effects of teachers' gender bias as measured by the IAT. The analysis sample consists of a matched teacher-student sample whose members participated in the survey data collection. The baseline controls in the first two columns include student gender, age, self-reported number of previous IAT solved, the IAT version solved (i.e., order compatible or incompatible), and teachers' gender, age, and children-bearing status. Observations are at the student level. I included grade and school fixed effects, and standard errors are clustered at the school level. Column (3) reports the effects of assessment-based gender biases using a matched dataset of students and teachers from administrative datasets. The controls include all controls from the main research design for identifying the impacts of grading bias on long-term outcomes. Observations are at the student-grade level. Standard errors are clustered at the school and student level in all columns. The students in both samples correspond to a young cohort with a projected graduation year from high school between 2020 and 2025.